\documentclass[%
twocolumn,
superscriptaddress,
showpacs,
amsmath,amssymb,
aps,pra,
floatfix
]{revtex4}
\usepackage{amsfonts}
\usepackage{amsmath}
\usepackage{graphicx}
\usepackage{bm}
\usepackage[usenames]{color}
\usepackage{mathtools}
\usepackage{verbatim}

\newcommand{\nt}{Nature}
\newcommand{\sci}{Science}

\newcommand{\njp}{New J. Phys.}

\begin{document}
\title{Triangular and Honeycomb Lattices of Cold Atoms in Optical Cavities}
\author{Shabnam Safaei}
\email{cqtshsa@nus.edu.sg}
\affiliation{Centre for Quantum Technologies, National University of Singapore, 3 Science Drive 2, 117543 Singapore}
\affiliation{MajuLab, CNRS-UNS-NUS-NTU International Joint Research Unit, UMI 3654, Singapore}
\author{Christian Miniatura}
\affiliation{MajuLab, CNRS-UNS-NUS-NTU International Joint Research Unit, UMI 3654, Singapore}
\affiliation{Centre for Quantum Technologies, National University of Singapore, 3 Science Drive 2, 117543 Singapore}
\affiliation{Physics Department, Faculty of Science, National University of Singapore, 2 Science Drive 3, 117551 Singapore}
\affiliation{INLN, Universit\'{e} de Nice-Sophia Antipolis, CNRS; 1361 route des Lucioles, 06560 Valbonne, France}
\affiliation{Institute of Advanced Studies, Nanyang Technological University, 60 Nanyang View, Singapore 639673, Singapore}
\author{Beno\^{\i}t Gr\'{e}maud}
\affiliation{MajuLab, CNRS-UNS-NUS-NTU International Joint Research Unit, UMI 3654, Singapore}
\affiliation{Centre for Quantum Technologies, National University of Singapore, 3 Science Drive 2, 117543 Singapore}
\affiliation{Physics Department, Faculty of Science, National University of Singapore, 2 Science Drive 3, 117551 Singapore}
\affiliation{Laboratoire Kastler Brossel, Ecole Normale Sup\'{e}rieure CNRS, UPMC; 4 Place Jussieu 75005 Paris, France}
\begin{abstract}
We consider a two-dimensional homogeneous ensemble of cold bosonic atoms loaded inside two optical cavities 
and pumped by a far-detuned external laser field. We examine the conditions for these atoms to self-organize 
into triangular and honeycomb lattices as a result of superradiance. By collectively scattering the pump photons, 
the atoms feed the initially empty cavity modes. As a result, the superposition of the pump and cavity fields 
creates a space-periodic light-shift external potential and atoms self-organize into the potential wells of 
this optical lattice. Depending on the phase of the cavity fields with respect to the pump laser, these 
minima can either form a triangular or a hexagonal lattice. By numerically solving the dynamical equations 
of the coupled atom-cavity system, we have shown that the two stable atomic structures at long times are the 
triangular lattice and the honeycomb lattice with equally-populated sites. We have also studied how to drive 
atoms from one lattice structure to another by dynamically changing the phase of the cavity fields with 
respect to the pump laser.
\end{abstract}
\maketitle
\section{Introduction}
\label{intro}

Cavity QED (CQED) investigates the interaction of atoms with confined electromagnetic field modes. When atoms are 
coupled to a high-finesse optical resonator, the usual free-space dipole force they experience is strongly 
enhanced and, at the same time, the back-action of atoms on the confined light field cannot be ignored any longer. 
As a consequence both atoms and light must be treated on the same footing and the dynamics for the atomic motion 
and the cavity field becomes strongly nonlinear. These hybrid systems open the way to new physical situations, in 
particular when cold atoms, bosonic or fermionic, are loaded inside optical cavities \cite{mekhov_quantum_2012,ritsch_cold_2013}. 
Furthermore, it is possible to probe the properties of these systems in a non-destructive way by using the field 
leaking outside the cavity \cite{mekhov_probing_2007}.

Trapping Bose-Einstein condensates (BEC) in laser-driven high-finesse optical cavities \cite{horak_coherent_2000} has been realized 
experimentally recently \cite{brennecke_cavity_2007,colombe_strong_2007,slama_superradiant_2007,wolke_cavity_2012}. The strong 
atom-cavity coupling enhances nonlinear effects and bistable behavior \cite{elsasser_optical_2004,gupta_cavity_2007,ritter_dynamical_2009,safaei_bistable_2013} 
and even chaos \cite{diver_nonlin_2014} set in. In the dispersive regime where the pump and cavity fields are both far-detuned from 
the atomic transition, the light fields impart forces on the atoms which thus move. In turn, the light fields pick up phase shifts 
induced by the refractive index of this moving atomic dielectric medium. This alters the light forces, thus the atomic motion, thus 
the accumulated dispersive phase shifts, and this combined atom-field process loops self-consistently. As a result, when the pump 
field strength is larger than some critical value, the atomic cloud scatters constructively the pump photons into the cavity modes. 
This causes an abrupt increase of the number of photons inside the cavity, a phenomenon referred to as superradiance, and the atoms 
achieve self-organization into the effective optical lattice created by the coherent superposition of the pump and cavity fields 
\cite{domokos_collective_2002,black_observation_2003,nagy_self-organization_2008,konya_multimode_2011,bux_cavity-controlled_2011}.
Self-organization breaks the initial translation symmetry. In other words, photon scattering couples the initial zero-momentum 
state of the atomic cloud to a superposition of higher recoil momentum states \cite{bhaseen_dynamics_2012}. This effect has been used 
to simulate the Dicke superradiance quantum phase transition \cite{dicke_coherence_1954,hepp_superradiant_1973,wang_phase_1973} 
in a BEC-cavity system where two collective motional modes of the condensate play the role of the two hyperfine spin states of the 
original Dicke model \cite{nagy_dicke-model_2010,baumann_dicke_2010,baumann_exploring_2011}. The non-equilibrium dynamics of such 
systems have been studied in \cite{nagy_critical_2011,konya_finite_2012,oztop_excitations_2012,bhaseen_dynamics_2012,torre_keldysh_2013,bakhtiari_noneq_2015}. 
Recent theoretical proposals have generalized the model even further by introducing cavity-assisted Raman coupling in order to reach 
phases such as self-organized magnetic lattices of bosons \cite{safaei_raman_2013} or topologically non-trivial phases of fermions \cite{pan_topolo_2015,kollath_ultracold_2015}.

In this article we consider a two-dimensional system similar to the experiment in \cite{baumann_dicke_2010} but with two crossing 
cavities in order to examine the possible formation of self-organized triangular and honeycomb lattices of cold bosons as a result 
of superradiance. Lately, the honeycomb lattice has attracted a lot of attention in the cold atom community because of its unique 
band structure mimicking Weyl-Dirac quasi-particles at the so-called Dirac points \cite{Lee2009, CastroNeto2009, Tarruell2012, Lim2012}. 
By numerically simulating the real-time dynamics of the cavity-atoms system we show that, depending on the relative phase $\phi$ of 
the cavity fields with respect to the pump field, three distinct atomic lattice structures are possible when non-interacting atoms 
and identical cavities are considered. The first one is a triangular lattice, the cavity fields being in phase with the pump laser 
($\phi= 0$). The second one is a honeycomb lattice with density-balanced sites, the cavities and pump fields being in quadrature 
($\phi= \pi/2$). The last one is a honeycomb lattice with density-imbalanced sites, the relative phase $\phi$ being in between zero 
and $\pi/2$. We address the long-time stability of these atomic lattices as well as the possibility of driving the atoms from one 
structure to another.

The rest of this paper is organized as follows. In Sec.~\ref{model} we introduce the experimental setup that we consider and give the 
relevant Hamiltonian in the dispersive regime (Sec.~\ref{hamiltonian}). The dynamical equations of the coupled atoms-cavities system 
are derived in Sec.~\ref{equations}. They give rise to an effective lattice potential for the center of mass motion of the atoms which 
is studied in details in Sec.~\ref{potential}, \ref{symmetry} and \ref{IdCav}. Using the symmetry properties of the effective lattice 
potential, we derive the atomic equations of motion in reciprocal space in Sec.~\ref{DMA}. The condition for the normal to superradiance 
phase transition is obtained by studying the linear response of the system in Sec.~\ref{instability}. Finally, in Sec.~\ref{numerics}, 
we present our numerical results about the atoms and cavity fields dynamics, the superradiance phase transition (Sec.~\ref{NtoS}), the 
long-time stability of the different atomic lattices which are obtained (Sec.~\ref{longtime}) as well as the possibility of switching 
the system between these different lattice structures (Sec.~\ref{switch}). We summarize the work in Sec.~\ref{sum}.

\section{Physical situation and model}
\label{model}

\begin{figure}[top]
\includegraphics[width=0.3\textwidth]{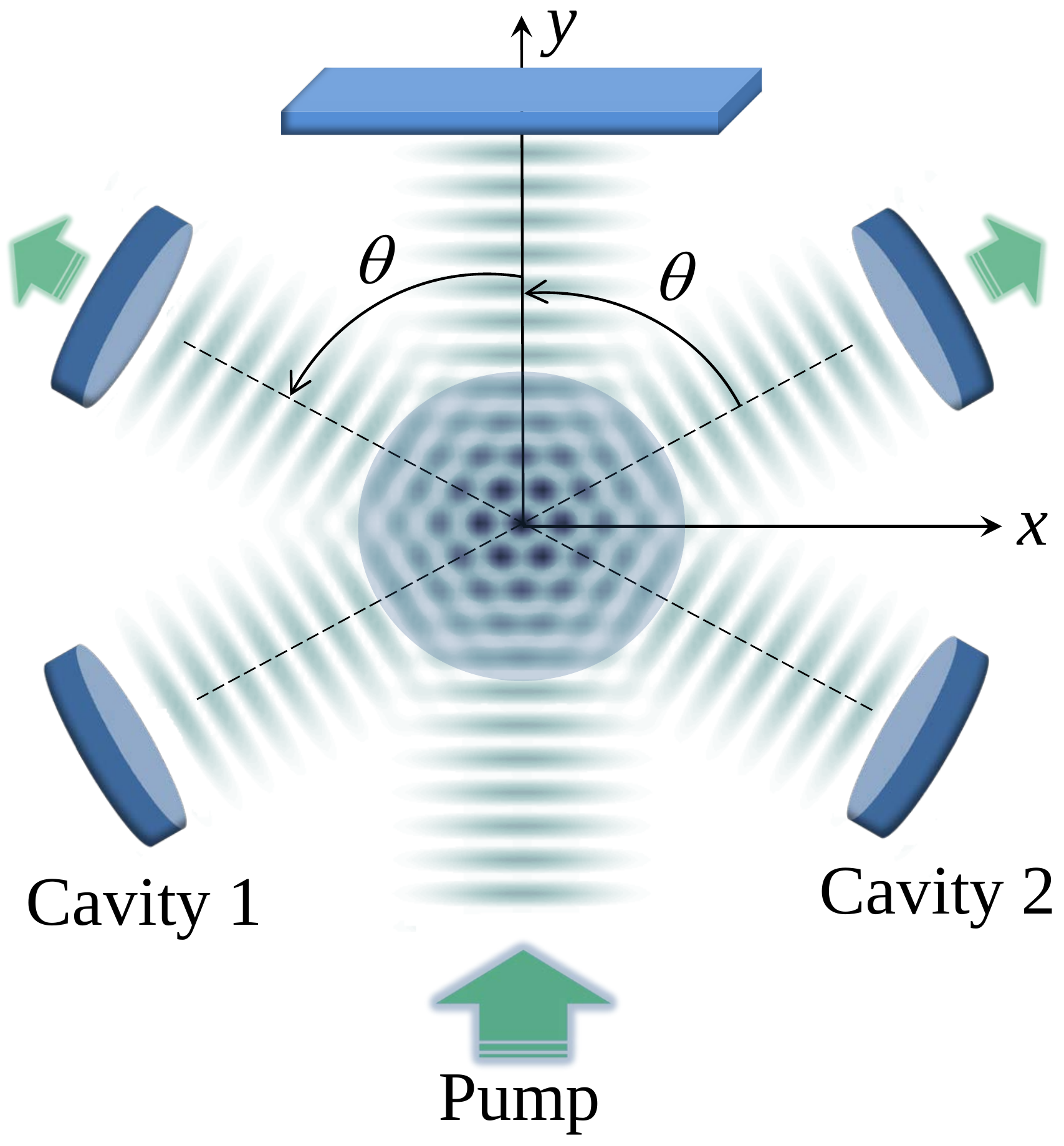}
\caption{(Color online) Schematic drawing of the setup that can be used to create 2D triangular 
and honeycomb lattices of bosonic cold atoms. A cloud of atoms (blue disk) is loaded inside two 
initially empty optical cavities symmetrically oriented by the angle $\theta = \pi/3$ about axis 
$Oy$. The atoms are then pumped by a retro-reflected classical laser field ($Oy$ direction) and 
release photons into the modes of the cavities. In return, the coherent superposition of the 
pump and cavity fields creates an effective periodic potential and atoms self-organize themselves 
into the corresponding potential wells. A triangular or honeycomb lattice structure of atoms thus 
emerges as a result of superradiance processes.
\label{fig:system}}
\end{figure}

\subsection{Hamiltonian}
\label{hamiltonian}

To study the self-organization of bosonic atoms into a triangular or a honeycomb lattice as a result of superradiance, we consider 
two high-finesse optical cavities with frequencies $\omega_j$ ($j=1,2$) located in the $xy$-plane and oriented symmetrically about 
the $y$-axis by an angle $\theta = \pi/3$ (see Fig.~\ref{fig:system}). 
A two-dimensional (2D) dilute cloud of non-interacting two-level ultracold bosonic atoms (resonance frequency $\omega_a$, excited-state 
lifetime $\Gamma$) is then loaded inside these cavities, which we assume do not contain any photons initially. A monochromatic 
standing-wave at frequency $\omega_p$ (obtained by retro-reflecting a classical laser field propagating along unit vector $\hat{{\bf y}}$) 
is used to pump the atoms. In such a system, the atoms scatter the pump field photons into the cavity modes and self-organize in the 
potential wells of the effective potential created by the standing-wave and cavity fields.

In the following, we assume that spontaneous emission processes are fully negligible so that the atomic dynamics is Hamiltonian. 
This is achieved when both the pump and cavity fields are far detuned from the atomic resonance frequency, $|\omega_p-\omega_a| \gg \Gamma$ 
and $|\omega_j-\omega_a| \gg \Gamma$. We further assume that the rotating-wave approximation is valid such that fast variables 
can be eliminated. This is the case when $|\omega_p-\omega_a| \ll \omega_p$ and $|\omega_p-\omega_j| \ll \omega_p$. Finally, we 
assume that one can adiabatically eliminate the atomic excited state amplitude so that atoms, initially prepared in their ground 
state, mostly evolve in their ground state. This is the case when the Rabi oscillation has a small amplitude, which happens when 
the pump detuning $\Delta_a =\omega_p-\omega_a$ is much larger in magnitude than the kinetic energy of the atoms in their excited 
state, the pump and cavity Rabi frequencies, and the pump-cavity detuning $\Delta_j =\omega_p-\omega_j$. Under all these assumptions, 
and since the pump is described by a classical field, the Hamiltonian describing the coupled dynamics between the cavity fields 
and the 2D motion of atoms, in their internal ground state and at center-of-mass position $\vec{r}=x\hat{{\bf x}}+y\hat{{\bf y}}$, reads:
\begin{eqnarray}
\label{eq:H}
H&=&\int d\vec{r}~\Psi^{\dag}{\cal{H}}\Psi
-\sum_{j=1,2}\hbar\Delta_{j} \,a^{\dag}_ja_j \\ 
\label{eq:calH1}
{\cal{H}}&=&
-\frac{\hbar^2}{2m}(\frac{\partial^2}{\partial x^2}+\frac{\partial^2}{\partial y^2})
+\frac{\hbar \Omega_p^2}{\Delta_a} F_p^2(\vec{r})
\nonumber\\
&+& \sum_{j=1,2}\frac{\hbar g_j^2}{\Delta_a} F^2_j(\vec{r})a^{\dag}_j a_j 
\nonumber\\
&+& \sum_{j=1,2}\frac{\hbar g_j\Omega_p }{\Delta_a} F_p(\vec{r})F_j(\vec{r})(a^{\dag}_j+a_j)
\nonumber\\
&+& \frac{\hbar g_1g_2}{\Delta_a} F_1(\vec{r})F_2(\vec{r})(a^{\dag}_1 a_2+ a^{\dag}_2 a_1),
\end{eqnarray}
where $\Omega_p$ is the pump Rabi frequency, $g_j$ the atom-cavity coupling strength, and $\Psi$ and $a_j$ the 
atomic and cavity bosonic annihilation operators. The atomic operator is normalized to the total number of atoms 
$N$, $\int d\vec{r} \, \Psi^\dag(\vec{r},t)\Psi(\vec{r},t) =N$.  With a suitable choice of the origin of coordinates, 
the pump and cavity mode functions can be written as 
\begin{eqnarray}
&&F_p(\vec{r}) = \cos(\vec{k}_p\cdot\vec{r}) \nonumber\\
&&F_j(\vec{r}) = \cos(\vec{k}_j\cdot\vec{r}+\phi/2), 
\end{eqnarray}
where $\phi$ is a controllable phase that can be changed, for example, by moving the pump mirror along $Oy$.
The pump and cavity wave vectors are $\vec{k}_p = k_p \hat{{\bf y}}$ and $\vec{k}_j = k_j \hat{{\bf b}}_j$ respectively, with the unit-length vectors
\begin{eqnarray}
\label{eq:RecVec}
\hat{{\bf b}}_1 &=& \sin\theta \, \hat{{\bf x}} + \cos\theta \, \hat{{\bf y}} = \frac{\sqrt{3}}{2}\hat{{\bf x}} + \frac{1}{2} \hat{{\bf y}} \nonumber \\
\hat{{\bf b}}_2 &=& -\sin\theta \, \hat{{\bf x}} + \cos\theta \, \hat{{\bf y}} = -\frac{\sqrt{3}}{2}\hat{{\bf x}} + \frac{1}{2} \hat{{\bf y}}.
\end{eqnarray}
The pump and cavity wavelengths are $\lambda_p=2\pi/k_p = 2\pi c/\omega_p$ and $\lambda_j=2\pi/k_j = 2\pi c/\omega_j$, $c$ being the speed of light.

\begin{figure}[top]
\includegraphics[width=0.22\textwidth]{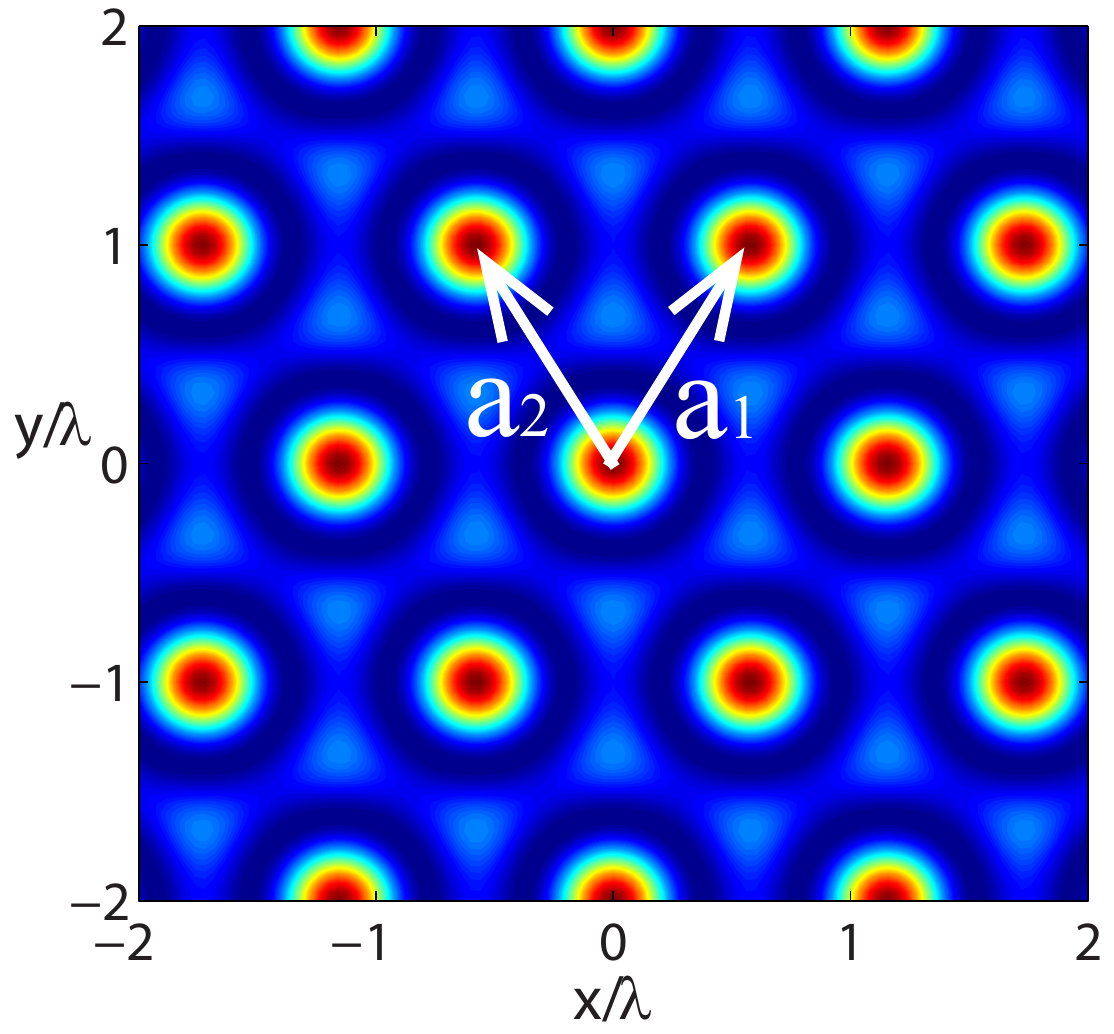}
\hspace{0.5cm}
\includegraphics[width=0.2\textwidth]{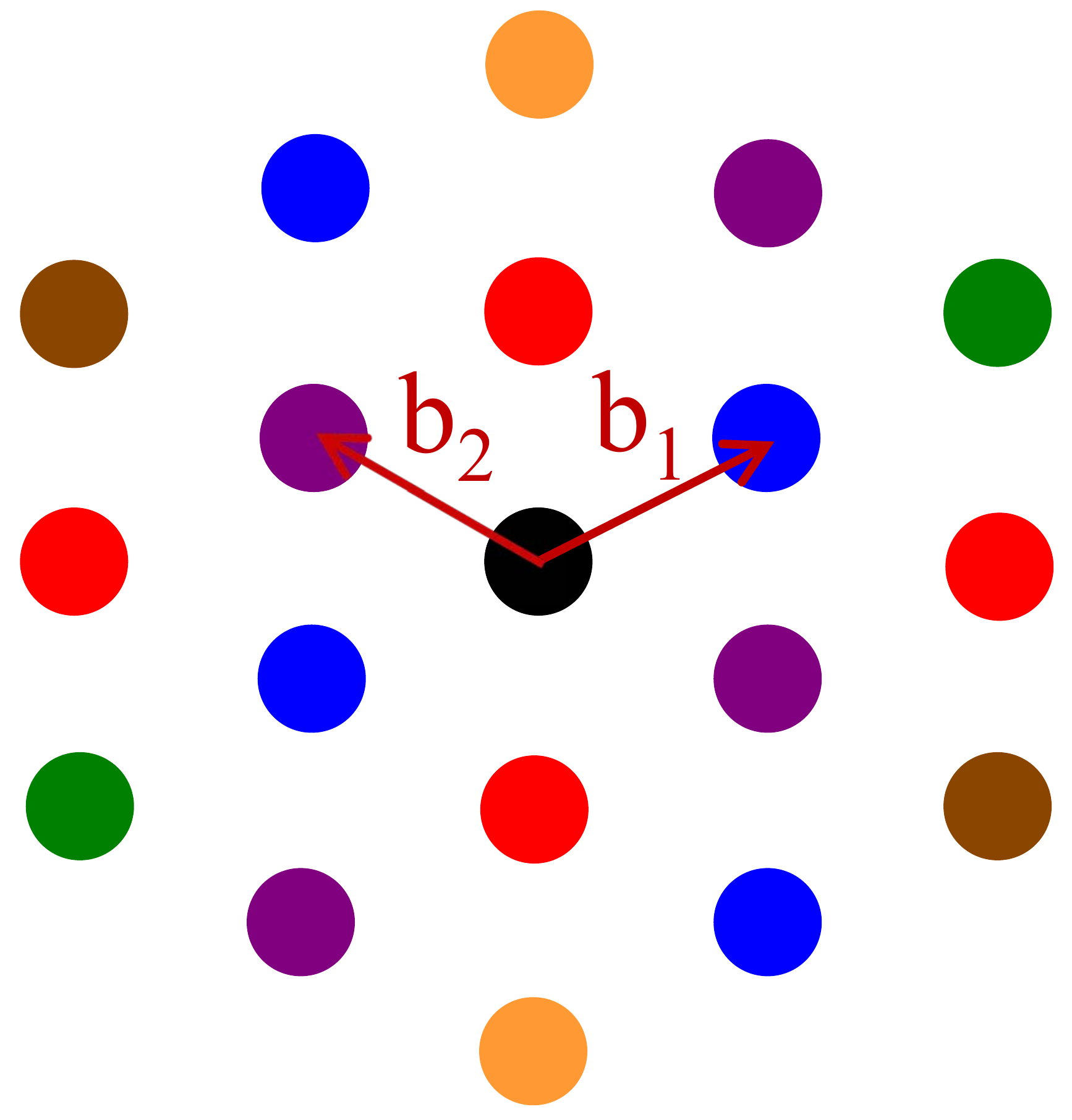}
\caption{(Color online) Left panel: Triangular Bravais lattice associated to the effective potential 
$V_{\text{eff}}(\vec{\rho}, \tau)$, Eq.~\eqref{eq:veff}, and its Bravais vectors $\vec{a}_1$ and 
$\vec{a}_2$, Eq.~\eqref{eq:BraVec}. Right panel: The corresponding triangular reciprocal lattice 
spanned by vectors $\hat{{\bf b}}_1$ and $\hat{{\bf b}}_2$, Eq.~\eqref{eq:RecVec}. The initial 
zero-momentum state of the atomic cloud (black disk at the center) is directly coupled to eighteen 
points of the reciprocal lattice (other colorful disks) by Schr\"odinger's equation Eq.~\eqref{eq:psidot}. 
Disks with the same color refer to a given coupling term in Schr\"odinger's equation Eq.~\eqref{eq:psidot}. 
Pump-cavity coupling: $\eta_1$ (purple) and $\eta_2$ (blue) terms. Inter-cavity coupling: $U_{12}$ term (red). 
Intra-cavity coupling: $U_1$ (green) and $U_2$ (brown) terms. Pump coupling: $U_p$ term (orange).
\label{fig:points}}
\end{figure}

\subsection{Dimensionless mean-field equations}
\label{equations}

In the rest of this paper, we consider the mean-field regime where the field operators are replaced by 
their mean values, $\Psi(\vec{r},t) \to \langle \Psi(\vec{r},t)\rangle \equiv \psi(\vec{r},t)$ and 
$a_j(t)\to \langle a_j(t)\rangle \equiv \alpha_j(t)$, and quantum fluctuations are discarded. We next 
define the pump recoil energy $E_R =  \hbar^2k_p^2/(2m)$. Using $k_p^{-1}$, $\omega^{-1}_R=\hbar/E_R$ 
and $E_R$ as space, time and energy units, the dimensionless Schr\"odinger equation for the atomic wavefunction reads:
\begin{eqnarray}
\label{eq:psidot}
&i \partial_\tau\varphi(\vec{\rho},\tau) =
\bigg[
-(\frac{\partial^2}{\partial \tilde{x}^2}+\frac{\partial^2}{\partial \tilde{y}^2})
+ U_p\cos^2\tilde{y} 
\nonumber\\
&+\sum_{j=1,2} \eta_j(\alpha_j+\alpha^*_j)\cos\tilde{y}\cos(\vec{b}'_j\cdot\vec{\rho}+\phi/2)
\nonumber\\
&+U_{12}(\alpha^*_1\alpha_2+\alpha_1\alpha^*_2)
\cos(\vec{b}'_1\cdot\vec{\rho}+\phi/2)\cos(\vec{b}'_2\cdot\vec{\rho}+\phi/2)
\nonumber\\
&+ \sum_{j=1,2} U_j |\alpha_j|^2\cos^2(\vec{b}'_j\cdot\vec{\rho}+\phi/2)
\bigg]\varphi(\vec{\rho},\tau)
\end{eqnarray}
where $\tau = \omega_Rt$, $\vec{\rho} = k_p\vec{r} = \tilde{x} \hat{{\bf x}}+\tilde{y}\hat{{\bf y}}$, 
$\varphi(\vec{\rho},\tau)=\psi(\vec{r},t)/k_p$, and $\vec{b}'_j = \vec{k}_j/k_p = (\lambda_p/\lambda_j) \hat{{\bf b}}_j$. 
Note that the normalization condition for the reduced atomic wavefunction is unchanged, $\int d\vec{\rho} \,|\varphi(\vec{\rho},\tau)|^2 = N$. 
The various dimensionless coupling constants appearing in Eq.~\eqref{eq:psidot} are:
\begin{equation}
\label{eq:CouPara}
U_p = \frac{\Omega_p^2}{\omega_R\Delta_a} \ U_j = \frac{g_j^2}{\omega_R\Delta_a} \
\eta_j =  \frac{g_j\Omega_p}{\omega_R\Delta_a} \
U_{12} = \frac{g_1g_2}{\omega_R\Delta_a}.
\end{equation}
Introducing the dimensionless cavity decay constants $\kappa_j$, the equations of motion for the cavity field amplitudes read:
\begin{eqnarray}
\label{eq:adot}
i\partial_\tau \alpha_1=
-(\delta_{1}+i\kappa_1)\alpha_1
+NU_{12}I_{12}\alpha_2
+N\eta_1I_{1p}
\nonumber\\
i\partial _\tau \alpha_2=
-(\delta_{2}+i\kappa_2)\alpha_2
+NU_{12}I_{12}\alpha_1
+N\eta_2I_{2p},
\end{eqnarray}
where $\delta_{j}=\Delta_{j}-NU_jI_j$ is the shifted cavity resonance frequency and where:
\begin{eqnarray}
\label{eq:intgs}
I_j &=&\frac{1}{N}
\int d\vec{\rho}~|\varphi(\vec{\rho},\tau)|^2\cos^2(\vec{b}'_j\cdot\vec{\rho}+\phi/2)
\nonumber\\
I_{12}&=&\frac{1}{N}
\int d\vec{\rho}~|\varphi(\vec{\rho},\tau)|^2\cos(\vec{b}'_1\cdot\vec{\rho}+\phi/2)\cos(\vec{b}'_2\cdot\vec{\rho}+\phi/2)
\nonumber\\
I_{jp}&=&\frac{1}{N}
\int d\vec{\rho}~|\varphi(\vec{\rho},\tau)|^2\cos\tilde{y}\cos(\vec{b}'_j\cdot\vec{\rho}+\phi/2).
\end{eqnarray}
Note that, because of the normalization of the atomic wave function, the $I$-integrals do not depend on the actual number of atoms $N$.

\subsection{Effective potential for atoms}
\label{potential} 

From Eq.~(\ref{eq:psidot}), one immediately sees that the pump and cavity fields create an effective potential for the atom center-of-mass dynamics:

\begin{eqnarray}
\label{eq:veff}
&V_{\text{eff}}(\vec{\rho},\tau)=
U_p\cos^2\tilde{y}+ \sum_{j=1,2} U_j |\alpha_j|^2\cos^2(\vec{b}'_j\cdot\vec{\rho}+\phi/2)
\nonumber\\
&+U_{12}(\alpha^*_1\alpha_2+\alpha^*_2\alpha_1)
\cos(\vec{b}'_1\cdot\vec{\rho}+\phi/2)\cos(\vec{b}'_2\cdot\vec{\rho}+\phi/2)
\nonumber\\
&+ \sum_{j=1,2} \eta_j (\alpha^*_j+\alpha_j)\cos \tilde{y}\cos(\vec{b}'_j\cdot\vec{\rho}+\phi/2).
\end{eqnarray}
This effective potential can be recast under the simpler and suggestive form 
\begin{equation}
\label{eq:veff1}
V_{\text{eff}}(\vec{\rho},\tau) = \frac{\hbar|\Omega(\vec{r},t)|^2}{E_R\Delta_a}, 
\end{equation}
featuring the complex Rabi field amplitude
 \begin{equation}
 \label{eq:Rabi}
 \Omega(\vec{r},t) = \Omega_p F_p(\vec{r}) + \sum_{j=1,2} g_j \alpha_j(t) F_j(\vec{r}) 
 \end{equation}
associated, in the mean-field regime, to the coherent superposition of the pump and cavity classical 
standing-wave electrical field amplitudes. As one can see, the pump and cavity fields are balanced when 
$n_j = |\alpha_j|^2 = (\Omega_p/g_j)^2= U_p/U_j$ ($j=1,2)$.

We will restrict our subsequent analysis to a red-detuned pump field, $\Delta_a<0$, for which the minima 
of the effective potential, Eqs.~\eqref{eq:veff}-\eqref{eq:veff1}, are found at the maxima of the norm of 
the Rabi field, Eq.~\eqref{eq:Rabi}, and will thus be the deepest. 

\subsection{Lattice symmetry of the effective potential}
\label{symmetry}

Let us define the two vectors $\vec{a}'_j= (\lambda_j/\lambda_p) \vec{a}_j$ ($j=1,2$) with
\begin{eqnarray}
\label{eq:BraVec}
\vec{a}_1 &=& \pi \, (\frac{\hat{{\bf x}}}{\sin\theta} + \frac{\hat{{\bf y}}}{\cos\theta})= 2\pi \, (\frac{\hat{{\bf x}}}{\sqrt{3}} + \hat{{\bf y}})\nonumber\\
\vec{a}_2 &=& \pi \, (-\frac{\hat{{\bf x}}}{\sin\theta} + \frac{\hat{{\bf y}}}{\cos\theta}) = 2\pi \, (-\frac{\hat{{\bf x}}}{\sqrt{3}} + \hat{{\bf y}})
\end{eqnarray}
and $|\vec{a}_j| = 4\pi/\sqrt{3}$. As readily checked, they satisfy the property $\vec{a}'_i\cdot\vec{b}'_j = 2\pi \delta_{ij}$. 
As a consequence, and by construction, the cavity mode functions $F_j$ are left invariant when $\vec{\rho}$ is translated by any 
integer combination of the $\vec{a}'_j$. The pump mode function $F_p$ is also left invariant provided $\vec{a}'_j.\hat{{\bf y}}$ 
is a multiple integer of $2\pi$. The simplest choice, which also fulfills the conditions leading to Eq.~\eqref{eq:calH1}, is 
$\lambda_j =\lambda_p$. If this condition is not strictly obeyed, $F_p$ will still be approximately invariant, to a very good 
accuracy, if $|k_p-k_j|L \ll 1$, where $L$ is the size of the atomic cloud. In the rest of the paper we will assume this 
condition to hold such that the approximation $\lambda_1 \approx \lambda_2 \approx \lambda_p \equiv \lambda$ is perfectly 
justified. In turn, it means that the cavity and pump fields have almost the same frequencies, $|\Delta_j |\ll \omega_p$~\footnote{Since $|\Delta_j |\ll \omega_p$, 
the cavity fields are also red-detuned from the atomic transition.}.

Within these assumptions, it follows that the Bravais lattice associated to the effective potential $V_{\text{eff}}(\vec{\rho},\tau)$ 
is $\mathcal{L}=\{\vec{R} = \sum_j l_j \vec{a}_j; l_j \in \mathbb{Z}\}$, see Fig.~\ref{fig:points}. It is triangular and its unit cell 
is $\mathcal{A} = \{\sum_i v_i\vec{a}_i;  0\leq v_i \leq 1 \}$ with dimensionless area $S= 2(2\pi)^2/\sqrt{3}$. The corresponding 
reciprocal lattice $\mathcal{R}=\{\vec{K} = \sum_j p_j \hat{{\bf b}}_j; p_j \in \mathbb{Z}\}$ is also triangular and its first Brillouin 
zone is $\mathcal{B} = \{\sum_ju_j\hat{{\bf b}}_j; |u_j| \leq 1/2 \}$ with dimensionless area $\Sigma = (2\pi)^2/S=\sqrt{3}/2$. 

\subsection{Case of identical cavities}
\label{IdCav}

In the following, we will examine the simplest case where the cavities have exactly the same characteristics, 
$\omega_1=\omega_2=\omega_c$, $g_1=g_2=g_c$ and $\kappa_1=\kappa_2=\kappa_c$. Then a reflection symmetry about axis 
$Oy$ ($x \to -x$) amounts to exchange the cavity fields, $\vec{k}_1 \leftrightarrow \vec{k}_2$. If we further assume 
that the cavities are initially empty and the initial atomic cloud is also symmetric with respect to $Oy$, then the 
subsequent time evolution will always enforce $\alpha_1=\alpha_2$ at all times~\footnote{As a word of caution, this 
is a mathematical statement. In real life, as is well-known in chaotic systems, numerical rounding-off errors may 
artificially break symmetries in actual simulations. An example is seen in Fig.~\ref{fig:pihalflong} where $n_1$ and $n_2$ are not perfectly equal.}. 
In this case the Rabi field, Eq.~\eqref{eq:Rabi}, writes:
\begin{equation}
\label{eq:DimRab}
\Omega(\vec{\rho},\tau) = \Omega_p \cos\tilde{y} + 2 \Omega_c(\tau) \cos(\frac{\tilde{y}+\phi}{2}) \cos(\frac{\sqrt{3}\tilde{x}}{2})
\end{equation}
with $\Omega_c(\tau)= g_c\alpha_1(\tau) = g_c\alpha_2(\tau)$. It is immediately seen that, in this case, the 
effective potential is always reflection-symmetric about planes located at $\tilde{x} = 2n\pi/\sqrt{3}$ ($n\in \mathbb{Z}$) 
and in particular about axis $Oy$ ($x\to -x$). It is also easy to prove that the potential for $\phi+ n\pi$ ($n\in\mathbb{Z}$) 
is obtained by a mere translation of the potential for $\phi$. Furthermore, the potential for $-\phi$ is simply 
obtained by reflecting the potential for $\phi$ about axis $Ox$ ($y\to-y$). This means that, for all practical purposes, 
the range of cavity phases can be restricted to $0\leq \phi \leq \pi/2$.

\begin{figure*}[top]
\includegraphics[width=0.3\textwidth]{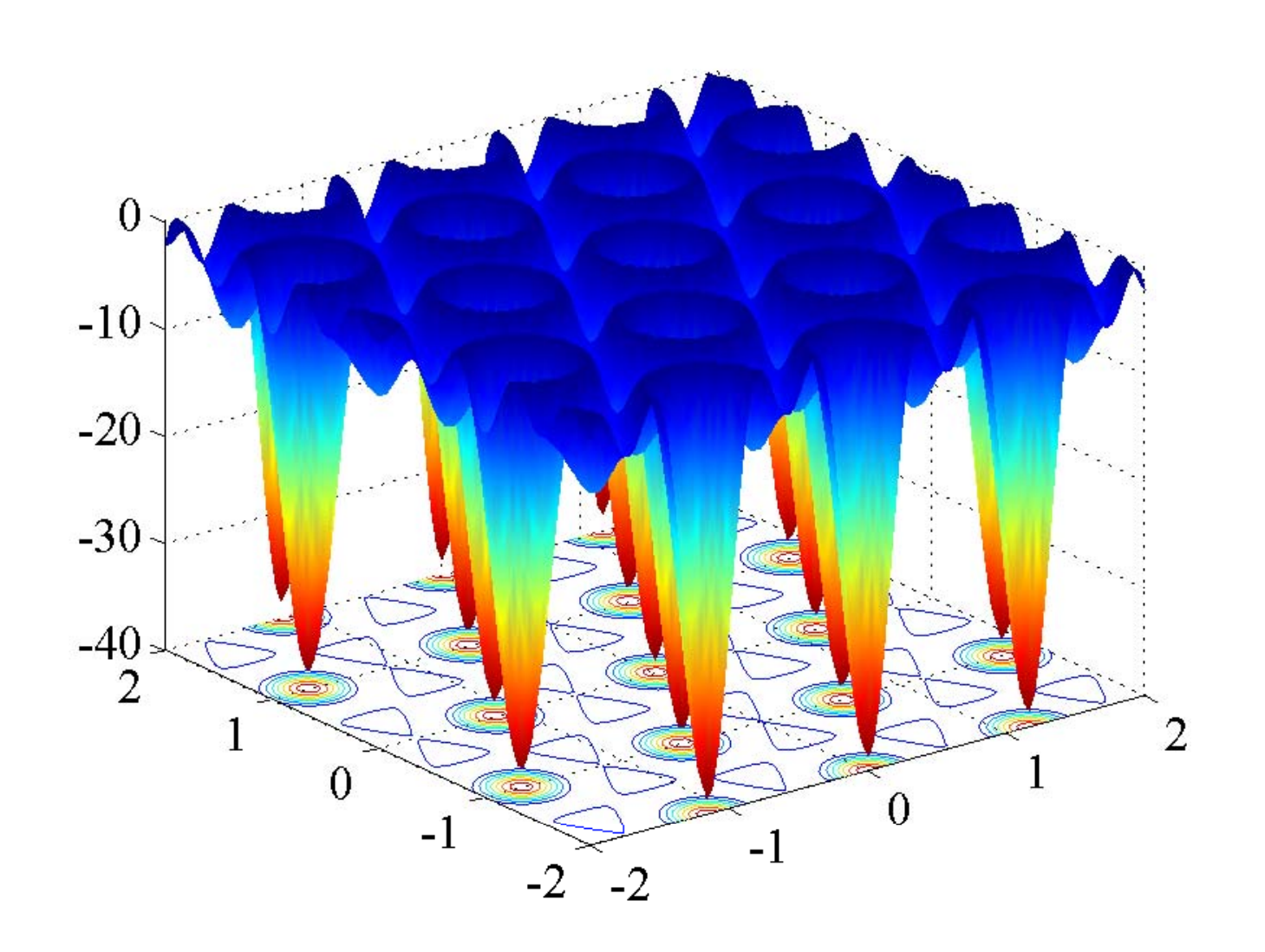}
\includegraphics[width=0.3\textwidth]{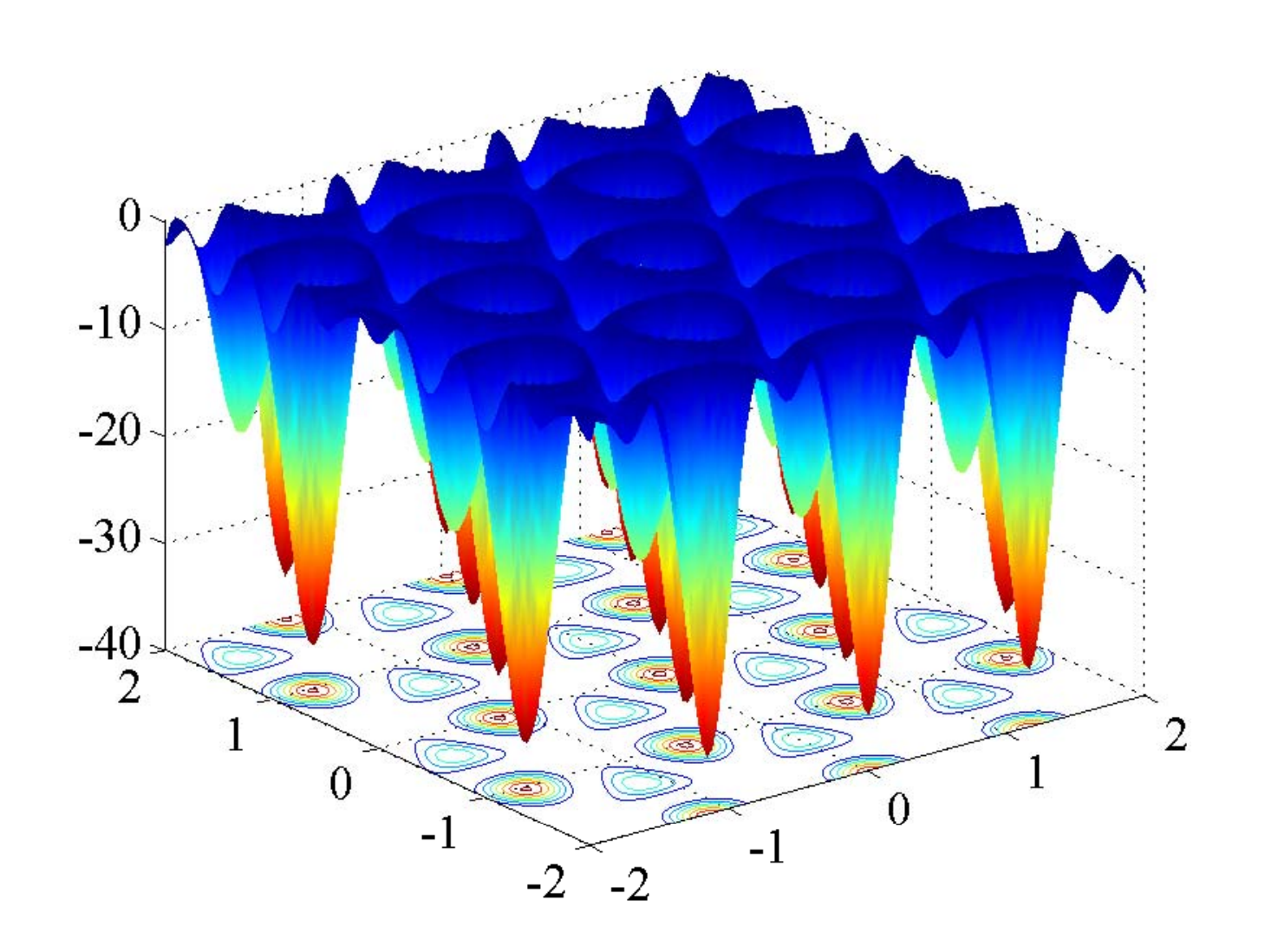}
\includegraphics[width=0.3\textwidth]{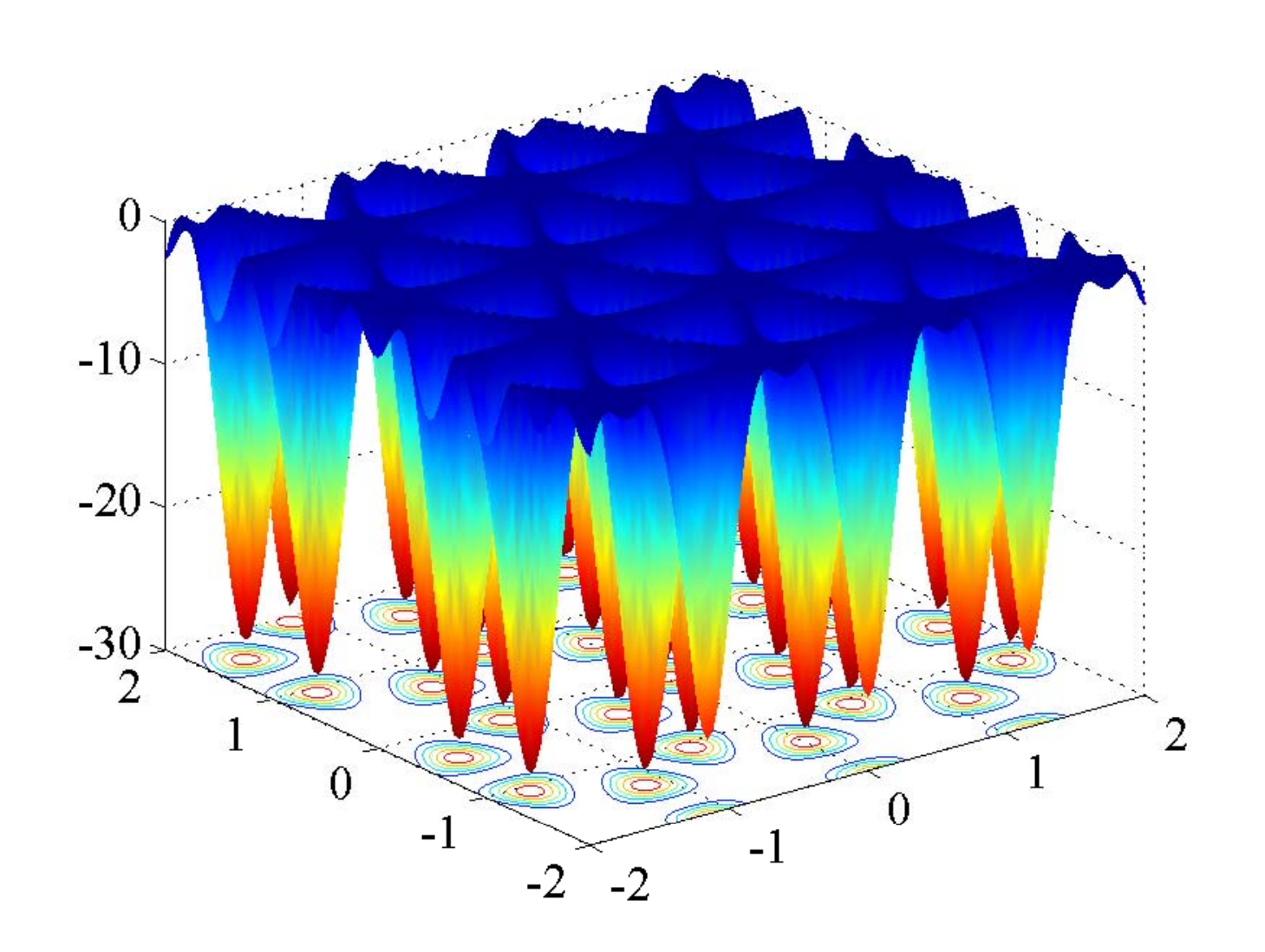}\\
\includegraphics[width=0.3\textwidth]{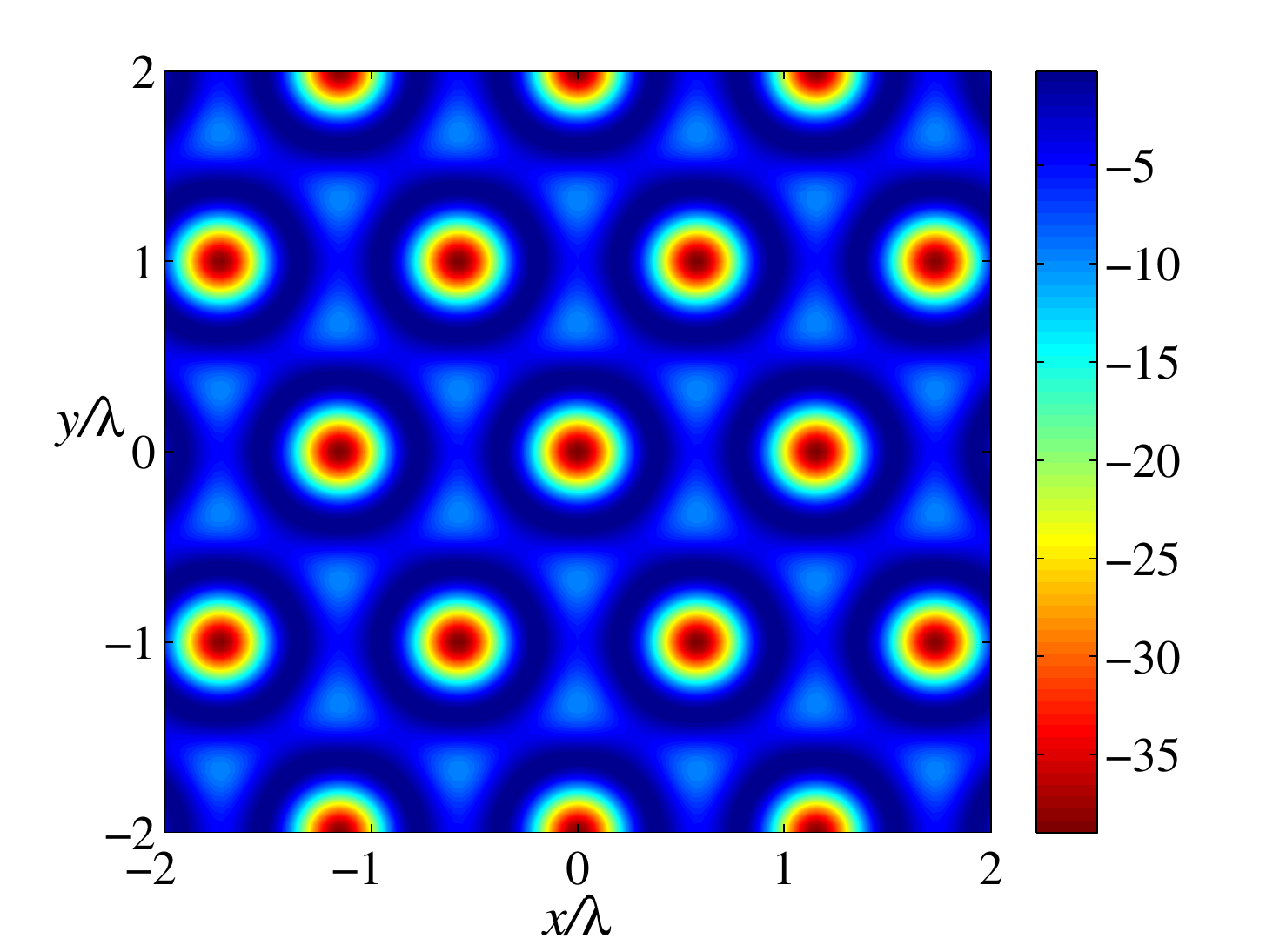}
\includegraphics[width=0.3\textwidth]{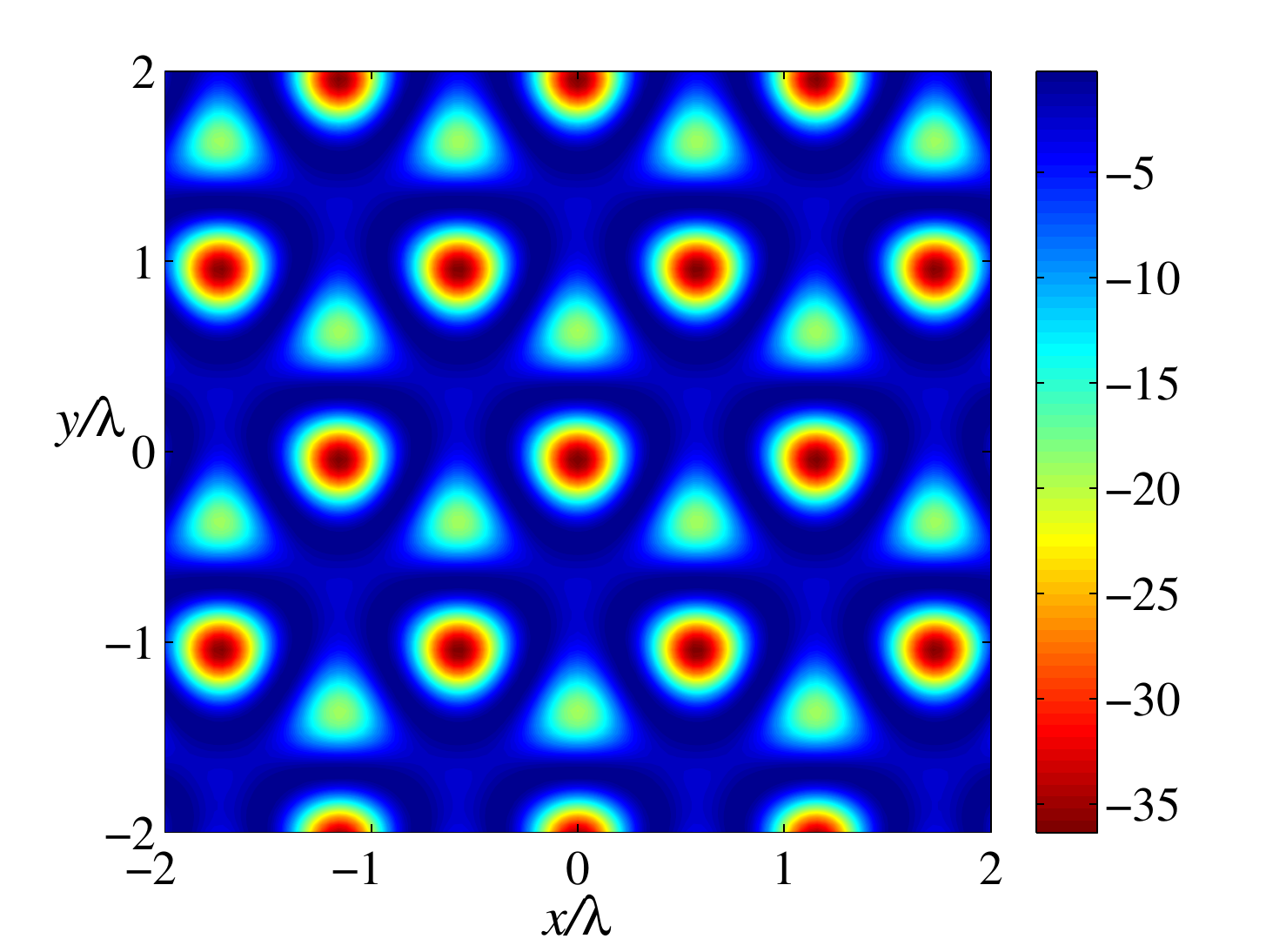}
\includegraphics[width=0.3\textwidth]{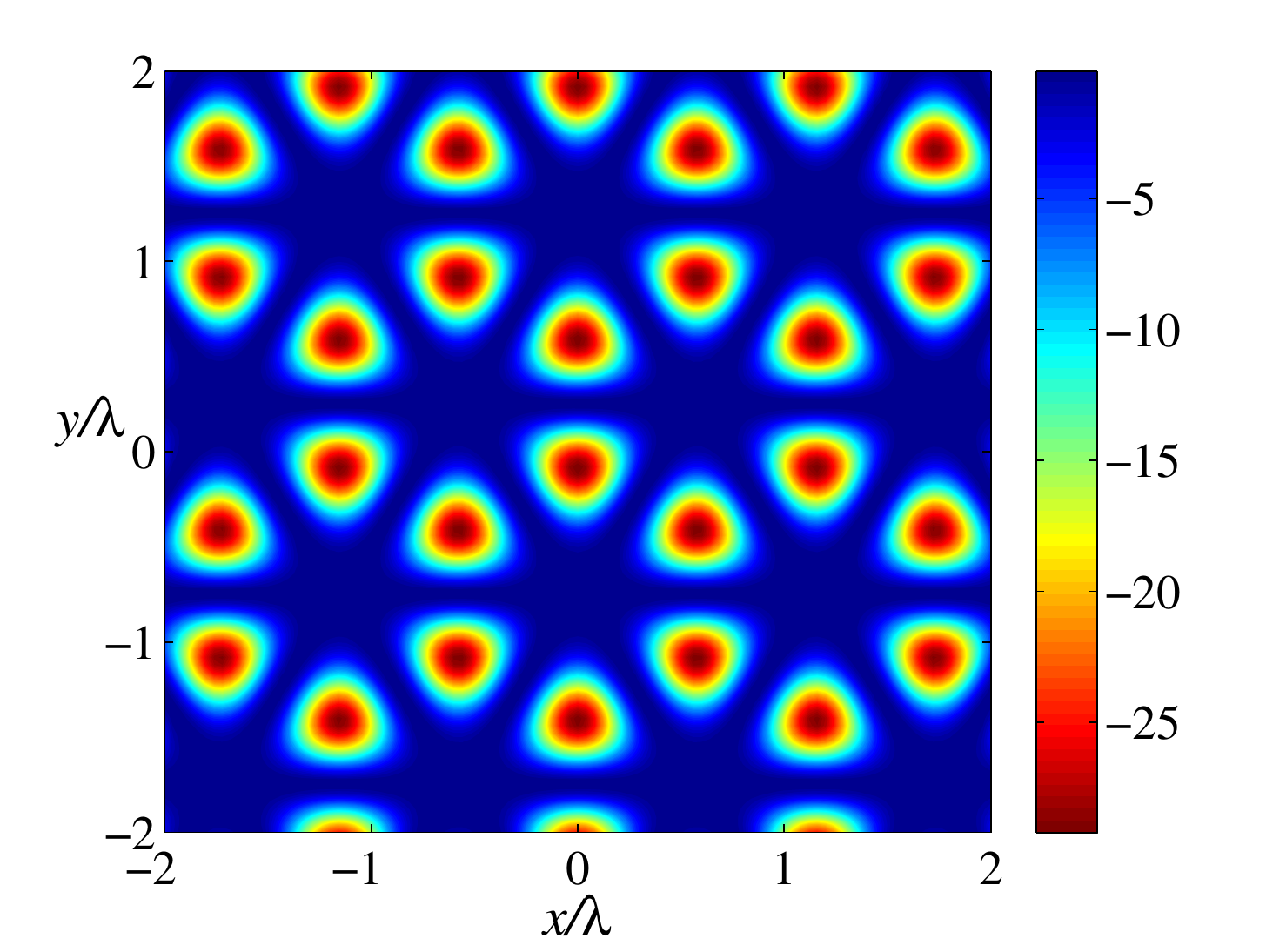}
\caption{(Color online) Effective potential $V_{\text{eff}}(\vec{\rho}, \tau)$ given by Eq.~(\ref{eq:veff}) and its 
lattice structure of minima (in units of the common wavelength $\lambda$ associated to the cavity and pump fields) 
for three values of the cavity phase $\phi$ appearing in the mode functions $F_j(\vec{r})$ (see text). Right panels: 
for $\phi=\pi/2$ a honeycomb structure is obtained for the potential minima which have all same depths. Middle panels: 
for lower values, e. g. $\phi=\pi/4$, a honeycomb structure is obtained but with energy-imbalanced minima. Left panels: 
for $\phi=0$, a triangular lattice is obtained for the potential minima. The dimensionless coupling parameters are 
$U_p=-4$, $U_j=U_{12}=-0.02$, $\eta_j = -\sqrt{2}/5$, see Eqs~\eqref{eq:CouPara}, and $\alpha_j(\tau)= \langle a_i(\tau)\rangle = 15$ ($j=1,2$). 
With these values, the pump and cavity fields are almost balanced since $U_p/U_j = 200$ and $n_j= |\alpha_j|^2=225$, see text after Eq.~\eqref{eq:Rabi}.
\label{fig:veff}
}
\end{figure*}

For $\phi=\pi/2$, the Bravais unit cell hosts two energy-balanced minima which can be labeled $A$ and $B$, so that 
the full lattice of minima is now a honeycomb lattice made of two shifted $A$ and $B$ triangular lattices, see 
Fig.~\ref{fig:veff}. Interestingly enough, this honeycomb structure does not depend on the relative weight between 
the pump and cavity Rabi amplitudes~\footnote{However nice-looking honeycomb structures are obtained with balanced 
or almost-balanced Rabi amplitudes.}. Indeed, when $\phi = \pi/2$, the potential is symmetric under $\vec{\rho} \to \pi\hat{{\bf y}}- \vec{\rho}$. 
This immediately ensures the existence of an even number of energy-balanced minima in the Bravais unit cell, here two.

When $\phi$ is continuously decreased from $\pi/2$, the $A$ and $B$ minima become energy-imbalanced, still 
retaining a nice-looking honeycomb structure when $\phi$ is not too small. This situation is interesting for 
producing bands with non-vanishing Chern numbers. When $\phi$ gets closer to zero, the honeycomb structure is 
lost for all practical purposes and the lattice of deepest minima is triangular.

Irrespective of the value of $\phi$, and for a priori different mean-field values $\alpha_j$, it is worth mentioning 
that the effective potential gets shifted by $\vec{a}_1/2$ when the sign of $\alpha_2$ is flipped, by $\vec{a}_2/2$ 
when the sign of $\alpha_1$ is flipped and by $(\vec{a}_1+\vec{a}_2)/2 = 2\pi \hat{{\bf y}}$ (which corresponds to a 
shift of $\lambda$ along axis $Oy$) when both signs are flipped. It is straightforward to check that Eqs~\eqref{eq:psidot} 
and \eqref{eq:adot} indeed remain invariant when the sign of $\alpha_1$ and/or $\alpha_2$ is flipped and the corresponding 
translation by $\vec{a}_j/2$ is implemented on the mode functions and atomic wave function $\varphi(\vec{\rho},\tau)$. 
Therefore, for each set of coupling parameters, there are four possible solutions associated to given numbers of cavity 
photons ($n_j=|\alpha_j|^2$), all related by translations along $\vec{a}_1/2$ and/or $\vec{a}_2/2$. Depending on the 
initial conditions, the system may select any one of these solutions. \\
%

\section{Schr\"odinger equation in reciprocal space}
\label{DMA}
We now rewrite Schr\"odinger's equation, Eq.~\eqref{eq:psidot}, by expanding the atomic wave function $\varphi(\vec{\rho},\tau)$ in reciprocal space:

\begin{eqnarray}
\label{eq:psimodes}
\varphi(\vec{\rho},\tau)=\sqrt{\frac{N_\mathcal{A}}{S}}
\sum_{\vec{K}\in\mathcal{R}} \sum_{\vec{q}\in\mathcal{B}} \, C_{\vec{K}}(\vec{q},\tau)e^{i(\vec{q}+\vec{K})\cdot\vec{\rho} }
\end{eqnarray}  
where $N_\mathcal{A}$ is the number of atoms in the unit Bravais cell $\mathcal{A}$.
The normalization conditions now read:
\begin{eqnarray}
&&\int_{\mathcal{A}} d\vec{\rho} \, |\varphi(\vec{\rho},\tau)|^2 = N_\mathcal{A} \\
&&\sum_{\vec{K}\in\mathcal{R}} \sum_{\vec{q}\in\mathcal{B}} \, |C_{\vec{K}}(\vec{q},\tau)|^2 = 1.
\end{eqnarray}

It is straightforward to see that $\vec{q}$ is actually conserved during the evolution. Indeed, the effective potential 
$V_{\text{eff}}(\vec{\rho},\tau)$ being periodic under Bravais translations, it cannot scatter and change $\vec{q}$. 
Substituting Eq.~\eqref{eq:psimodes} into Eq.~\eqref{eq:psidot} and using Eq.~\eqref{eq:RecVec}, we obtain the following 
dynamical equations for the coefficients $C_{\vec{K}}(\vec{q},\tau)$:
\begin{widetext}
\begin{eqnarray}
\label{eq:cdot}
i\partial_\tau C_{\vec{K}}&=&
(
\vec{q}+\vec{K}
)^2 C_{\vec{K}}
+ \frac{U_p}{4}
\Big(
2C_{\vec{K}}+C_{\vec{K}+2\hat{{\bf b}}_1 +2 \hat{{\bf b}}_2}+C_{\vec{K}-2\hat{{\bf b}}_1 -2 \hat{{\bf b}}_2}\Big)  + \sum_{j=1,2} \frac{U_j|\alpha_j|^2}{4}
\Big(
2C_{\vec{K}}+e^{i\phi}C_{\vec{K}-2\hat{{\bf b}}_j}+e^{-i\phi}C_{\vec{K}+2\hat{{\bf b}}_j}
\Big)\nonumber\\
&+&\sum_{j=1,2}\frac{\eta_j(\alpha_j+\alpha^*_j)}{4}
\Big(
e^{i\phi/2}C_{\vec{K}-\hat{{\bf b}}_1-\hat{{\bf b}}_2-\hat{{\bf b}}_j}+e^{-i\phi/2}C_{\vec{K}+\hat{{\bf b}}_1+\hat{{\bf b}}_2+\hat{{\bf b}}_j}+
e^{i\phi/2}C_{\vec{K}+\hat{{\bf b}}_1+\hat{{\bf b}}_2-\hat{{\bf b}}_j}+e^{-i\phi/2}C_{\vec{K}-\hat{{\bf b}}_1-\hat{{\bf b}}_2+\hat{{\bf b}}_j}
\Big) \nonumber\\
&+&\frac{U_{12}(\alpha_1^*\alpha_2+\alpha_2^*\alpha_1)}{4}
\Big(
e^{i\phi}C_{\vec{K}-\hat{{\bf b}}_1-\hat{{\bf b}}_2}+
e^{-i\phi}C_{\vec{K}+\hat{{\bf b}}_1+\hat{{\bf b}}_2}+
C_{\vec{K}-\hat{{\bf b}}_1+\hat{{\bf b}}_2}+
C_{\vec{K}+\hat{{\bf b}}_1-\hat{{\bf b}}_2}
\Big)
\end{eqnarray}
\end{widetext}
where the $\vec{q}$-dependence of the $C$-coefficients is the same on both sides of the equation. The dynamical 
equations for the cavity fields $\alpha_j$, Eq.~\eqref{eq:adot}, remain the same but with the integrals in 
Eq.~\eqref{eq:intgs} now reading: 
\begin{widetext}
\begin{eqnarray}
\label{eq:intgs2}
&&I_j=\frac{1}{4}\sum_{\vec{K} \vec{q}}
C^*_{\vec{K}}\Big(
2C_{\vec{K}}+
e^{-i\phi}C_{\vec{K}+2\hat{{\bf b}}_j}+
e^{i\phi}C_{\vec{K}-2\hat{{\bf b}}_j}
\Big)
\\
&&I_{12}=\frac{1}{4}\sum_{\vec{K},\vec{q}}
C^*_{\vec{K}}\Big(
e^{-i\phi}C_{\vec{K}+\hat{{\bf b}}_1+\hat{{\bf b}}_2}+
e^{i\phi}C_{\vec{K}-\hat{{\bf b}}_1-\hat{{\bf b}}_2}+
C_{\vec{K}+\hat{{\bf b}}_1-\hat{{\bf b}}_2}+
C_{\vec{K}-\hat{{\bf b}}_1+\hat{{\bf b}}_2}
\Big)\nonumber\\
&&I_{jp}=\frac{1}{4}\sum_{\vec{K},\vec{q}}
C^*_{\vec{K}}\Big(
e^{-i\phi/2}C_{\vec{K}+\hat{{\bf b}}_1+\hat{{\bf b}}_2+\hat{{\bf b}}_j}+
e^{i\phi/2}C_{\vec{K}-\hat{{\bf b}}_1-\hat{{\bf b}}_2-\hat{{\bf b}}_j}+
e^{-i\phi/2}C_{\vec{K}-\hat{{\bf b}}_1-\hat{{\bf b}}_2+\hat{{\bf b}}_j}+
e^{i\phi/2}C_{\vec{K}+\hat{{\bf b}}_1+\hat{{\bf b}}_2-\hat{{\bf b}}_j}
\Big)
\nonumber
\end{eqnarray}
\end{widetext}

Generally, in superradiance and self-organization problems, the atomic system is initially homogeneous (zero-momentum state), 
meaning that only the $C$-coefficient with $\vec{K}=0$ and $\vec{q}=0$ is nonzero initially (and equal to 1 because of the 
normalization condition). In the course of time, as a result of superradiance, higher-momentum states will be occupied. For 
our system, the dynamical equations (\ref{eq:cdot}) couple $C_{\vec{K}=0}(\vec{q}=0,t=0)$ to the eighteen points of the 
reciprocal lattice $\mathcal{R}$ shown in Fig.~\ref{fig:points}. Each of these points will be in turn coupled to other points 
in the reciprocal lattice with even higher momenta. In the following sections, we restrict our calculations 
to momenta transfers less than $3\hbar k$, i.e. to reciprocal lattice points $\vec{K} = \sum_j p_j \hat{{\bf b}}_j$ with 
$|p_j| \leq 3$ and $|p_1-p_2|\le 3$. These thirty-seven points in total include the initial zero-momentum state, the 
eighteen states to which this state is directly coupled, and the eighteen other states to which the states with momentum 
$\hbar k$ (the points closest to the initial zero-momentum state) are directly coupled. 

\section{Instability and Phase Transition Condition}
\label{instability}
\begin{figure*}[top]
\includegraphics[width=0.43\textwidth]{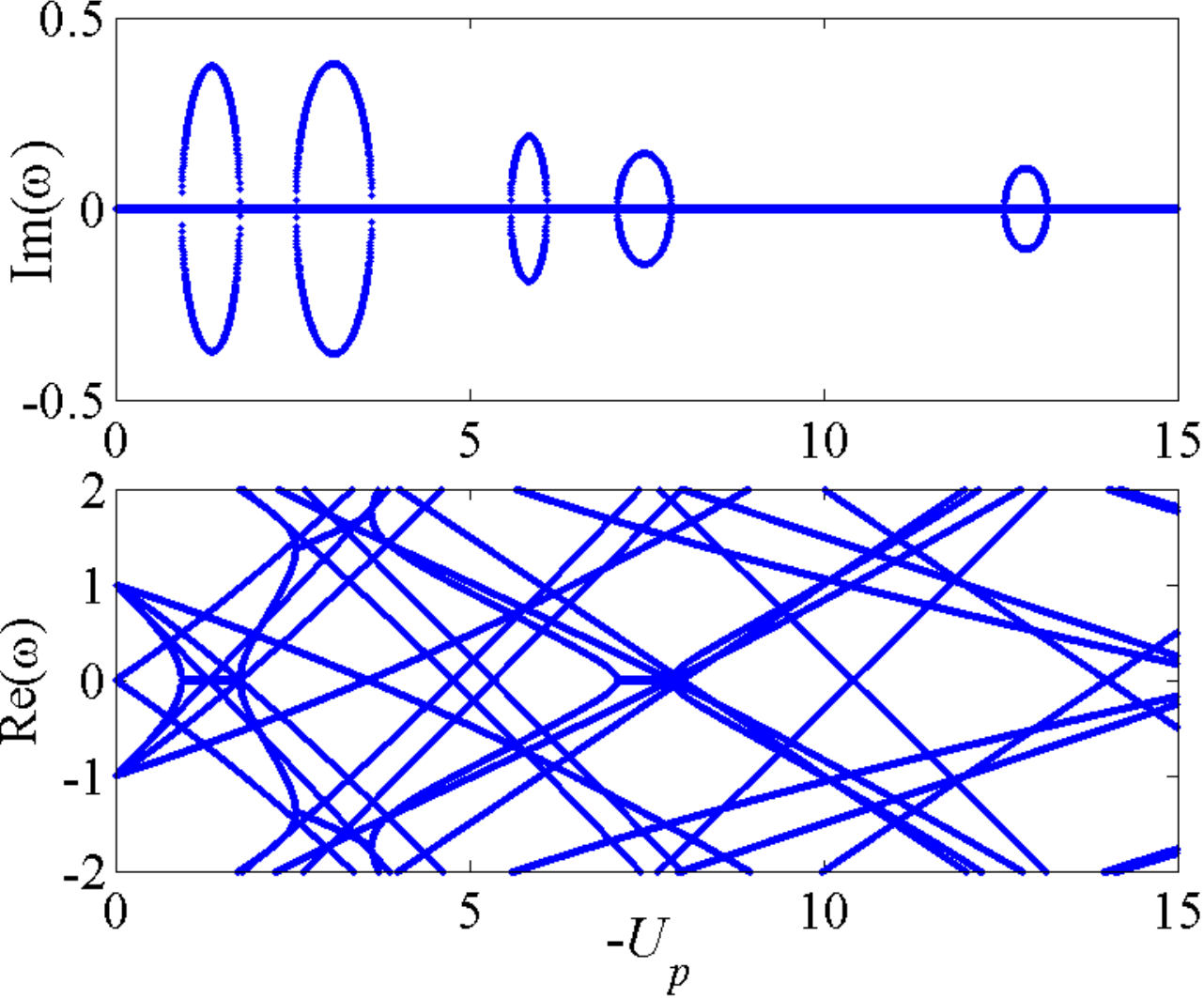}
\hspace{1cm}
\includegraphics[width=0.43\textwidth]{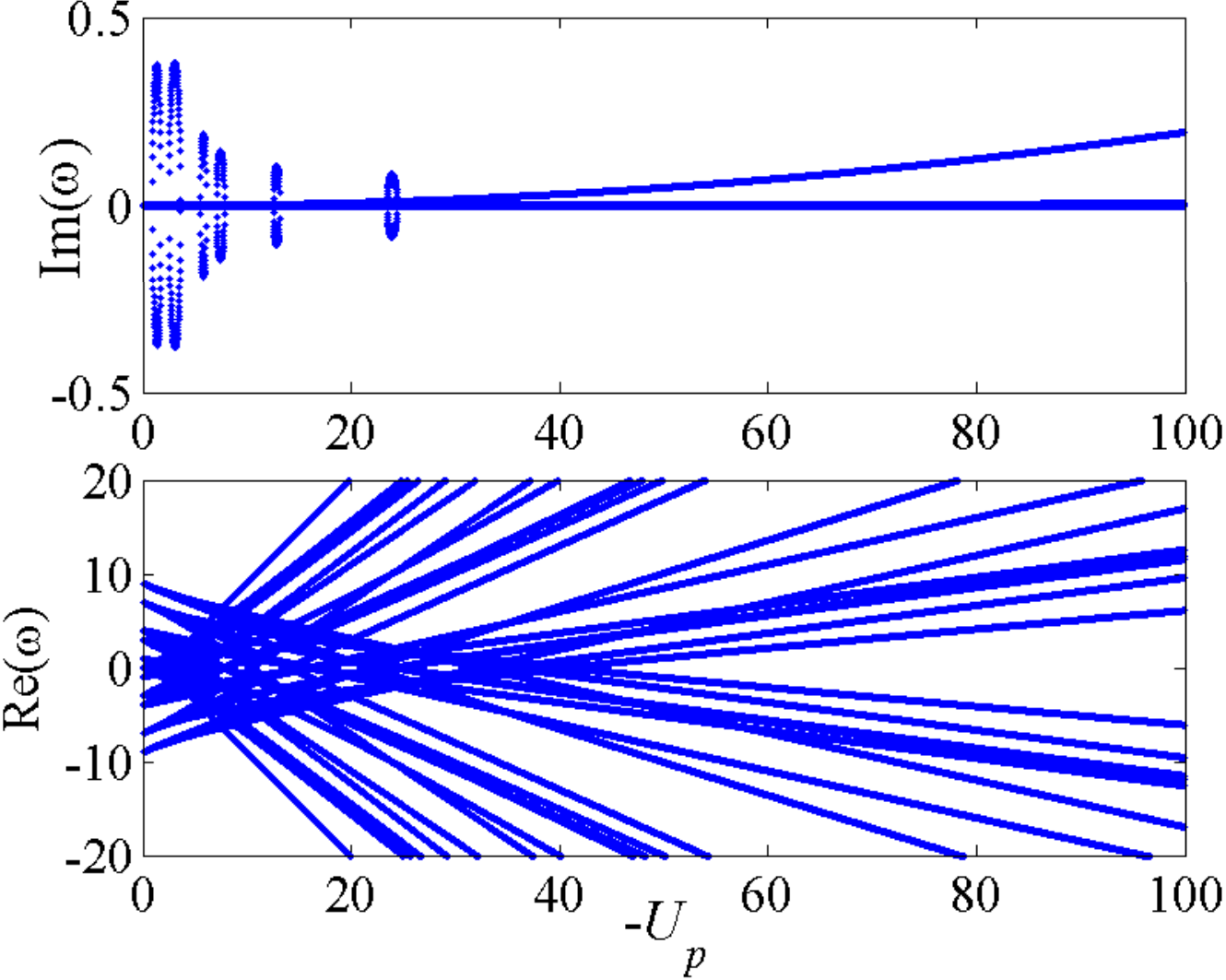}
\caption{Imaginary (top panels) and real parts (bottom panels) of the eigenvalues $\omega$ of 
stability matrix $\mathcal{M}$ as functions of the (negative) dimensionless pump coupling parameter $U_p$ (red-detuned pump). 
We have used thirty-seven reciprocal lattice vectors to compute and diagonalize $\mathcal{M}$, see discussion end of Sec.~\ref{DMA}. 
Positive imaginary parts signal instability of our initial state (empty cavities, homogeneous atomic cloud). The dimensionless 
cavity parameters are $U_j=U_{12}=-1.74\times 10^{-2}$, $\Delta_j=-4.7\times 10^3$ and $\kappa_j=480$ ($j=1,2$). The total 
number of atoms is $N=1.5\times 10^5$.
\label{fig:omega}}
\end{figure*}

In this section, we study the conditions for superradiance to take place and focus on the pump strength needed to destabilize 
the initial state for given cavity parameters. For this purpose we consider the linear response of the system to perturbations 
added to the initial cavity fields and atomic wave function. We thus substitute $\alpha^{(0)}_j\rightarrow\alpha^{(0)}_j+\delta\alpha_j$ 
and $C^{(0)}_{\vec{K}}\rightarrow C^{(0)}_{\vec{K}}+\delta C_{\vec{K}}$ into the dynamical equations Eqs~\eqref{eq:adot} and \eqref{eq:cdot} 
and in the integrals given in Eq.~\eqref{eq:intgs2}. Our initial state is $\alpha^{(0)}_j =0$ (empty cavities) and 
$C^{(0)}_{\vec{K}}  = \delta_{\vec{K}0}\delta_{\vec{q}0}$ (homogeneous cloud). 
For this initial state, $I_j = 1/2$ and $I_{12} =0$. Since $\vec{q}$ cannot change under evolution, as explained in Sec.~\ref{DMA}, 
its value remains zero and will thus be omitted. The linearized perturbed equations then read:
\begin{eqnarray}
\label{eq:deladot}
i\partial_\tau\delta\alpha_j =-\Big(\Delta_{j}-\frac{NU_j}{2}+i\kappa_j\Big)\delta\alpha_j+N\eta_j\delta I_{jp},
\end{eqnarray}
with
\begin{widetext}
\begin{eqnarray}
\label{eq:delIl}
\delta I_{jp}&=& \frac{1}{4}\sum_{\vec{K}} C^{(0)*}_{\vec{K}} \Big(
e^{-i\phi/2}\delta C_{\vec{K}+\hat{{\bf b}}_1+\hat{{\bf b}}_2+\hat{{\bf b}}_j}+
e^{i\phi/2} \delta C_{\vec{K}-\hat{{\bf b}}_1-\hat{{\bf b}}_2-\hat{{\bf b}}_j}+
e^{-i\phi/2} \delta C_{\vec{K}-\hat{{\bf b}}_1-\hat{{\bf b}}_2+\hat{{\bf b}}_j}+
e^{i\phi/2} \delta C_{\vec{K}+\hat{{\bf b}}_1+\hat{{\bf b}}_2-\hat{{\bf b}}_j}
\Big) \nonumber \\
&+&\frac{1}{4}\sum_{\vec{K}} \delta C^*_{\vec{K}}\Big(
e^{-i\phi/2}C^{(0)}_{\vec{K}+\hat{{\bf b}}_1+\hat{{\bf b}}_2+\hat{{\bf b}}_j}+
e^{i\phi/2}C^{(0)}_{\vec{K}-\hat{{\bf b}}_1-\hat{{\bf b}}_2-\hat{{\bf b}}_j}+
e^{-i\phi/2}C^{(0)}_{\vec{K}-\hat{{\bf b}}_1-\hat{{\bf b}}_2+\hat{{\bf b}}_j}+
e^{i\phi/2}C^{(0)}_{\vec{K}+\hat{{\bf b}}_1+\hat{{\bf b}}_2-\hat{{\bf b}}_j}
\Big).
\end{eqnarray}
and
\begin{eqnarray}
i\partial_\tau\delta C_{\vec{K}}
&=&(\vec{K}^2 + \frac{U_p}{2}) \delta C_{\vec{K}}
+\frac{U_p}{4} \Big(\delta C_{\vec{K}+2\hat{{\bf b}}_1 +2 \hat{{\bf b}}_2}+\delta C_{\vec{K}-2\hat{{\bf b}}_1 -2 \hat{{\bf b}}_2}
\Big)+\sum_{j=1,2}\frac{\eta_j(\delta\alpha_j+\delta\alpha^*_j)}{4}
\Big(
e^{i\phi/2}C^{(0)}_{\vec{K}-\hat{{\bf b}}_1-\hat{{\bf b}}_2-\hat{{\bf b}}_j} \nonumber\\
&+&e^{-i\phi/2}C^{(0)}_{\vec{K}+\hat{{\bf b}}_1+\hat{{\bf b}}_2+\hat{{\bf b}}_j} 
+ \, e^{i\phi/2}C^{(0)}_{\vec{K}+\hat{{\bf b}}_1+\hat{{\bf b}}_2-\hat{{\bf b}}_j}+e^{-i\phi/2}C^{(0)}_{\vec{K}-\hat{{\bf b}}_1-\hat{{\bf b}}_2+\hat{{\bf b}}_j}
\Big)  \label{eq:delcdot}
\end{eqnarray}
\end{widetext}
Eqs~\eqref{eq:delIl} and \eqref{eq:delcdot} only involve reciprocal lattice vectors of the form ${\vec K}+\vec{Q}$ with a transfer lattice vector 
$\vec{Q} = \sum_j Q_j \hat{{\bf b}}_j$ satisfying $Q_j= 0, \pm 1, \pm 2$ and $(Q_1-Q_2) = 0, \pm1$. Then it can be seen that all phase factors can be gauged away: 
\begin{eqnarray}
&&C^{(0)}_{\vec{K}+\vec{Q}} \to \tilde{C}^{(0)}_{\vec{K}+\vec{Q}} = e^{i r(\vec{Q}) \phi/2} \, C^{(0)}_{\vec{K}+\vec{Q}} \\
&&\delta C_{\vec{K}+\vec{Q}} \to \delta \tilde{C}_{\vec{K}+\vec{Q}} = e^{i r(\vec{Q}) \phi/2} \, \delta C_{\vec{K}+\vec{Q}},
\end{eqnarray}
where $(Q_1+Q_2)=4n+r(\vec{Q})$, with $n\in\mathbb{Z}$, $|r(\vec{Q})| \leq 3$ and $r(\vec{Q})$ the (positive or negative) remainder with smallest absolute value.
In other words, the stability of our initial state is independent from the cavity phase which can thus be set to $\phi=0$ in the stability analysis.

Let us define the column vector $\underline{Y} = (\{\delta C_{\vec{K}}\}, \{\delta\alpha_j\})^T$, where the superscript $T$ denotes transposition. 
The linearized equations Eqs~\eqref{eq:deladot} and \eqref{eq:delcdot} can then be rewritten in compact form $i\partial_\tau \underline{X}=\mathcal{M}\underline{X}$ 
where the stability matrix $\mathcal{M}$ controls the dynamics of the (doubled) perturbation column vector $\underline{X}=(\underline{Y}, \underline{Y}^*)^T$ 
around our initial condition. As one can see, the perturbation will grow, and the initial system is unstable, as soon as the eigenvalues $\omega$ of $\mathcal{M}$ 
have a positive imaginary part. The matrix $\mathcal{M}$ has the following 4-block structure
\begin{equation}
\mathcal{M}=
  \begin{bmatrix}
    F & G \\
    -G^* & -F^*
  \end{bmatrix}
 \end{equation}
where the star stands for complex conjugation and where the sub-matrix $G$ is symmetric $G^T=G$. It is easy to show that 
$\sigma_x\mathcal{M}\sigma_x=-\mathcal{M}^*$ , where the usual entries of the Pauli matrix $\sigma_x$ have been replaced by 
the null and identity matrices. As a consequence, $\omega$ and $-\omega^*$ are both eigenvalues of $\mathcal{M}$, with 
eigenstates $\underline{X}_\omega$ and $\sigma_x\underline{X}^*_\omega$, and the real parts must come in opposite pairs.

Fig.~\ref{fig:omega} shows the real and imaginary parts of the eigenvalues of the stability matrix $\mathcal{M}$ as functions 
of the pump coupling parameter $U_p$ and obtained
for fixed typical cavity experimental parameters. Note that we have not reproduced in these plots eigenvalues with large 
imaginary parts (equal to the cavity damping rates $\kappa_j$). As expected from the symmetry of $\mathcal{M}$, one can see that the real parts 
come indeed in opposite pairs. However, contrary to what a quick and misleading glance at the plots may let think, imaginary parts should not 
come in opposite pairs. Indeed, the sub-matrix $F$ becomes Hermitian, $F^\dag = F$, only when the cavities are lossless which implies the 
additional symmetry $\sigma_z\mathcal{M}\sigma_z=\mathcal{M}^\dag$. In this particular case, both $\omega$ and $\omega^*$ are eigenvalues 
and the imaginary parts would also come in opposite pairs. This is not the case here since the cavities are lossy ($\kappa_j \neq 0$). 
And indeed, a careful check shows that seemingly opposite values of the imaginary parts are in fact slightly different. 
We have checked that this difference approaches zero when the cavity damping rates $\kappa_j$ go to zero and the cavities become lossless. 
Another important point to mention is that all the imaginary parts (positive or negative) are either two- 
or four-fold degenerate. The two-fold degeneracy is always present and comes from the fact that $\omega$ and $-\omega^*$ are both eigenvalues. 
The occasional four-fold degeneracy comes from an additional spatial symmetry of the system: when cavities are identical, the system is 
reflection-symmetric about axis $Oy$ regardless of the value of $\phi$, see discussion after Eq.~\eqref{eq:DimRab}. Therefore, in some cases, 
there are two additional eigenstates of $\mathcal{M}$ co-existing with $\underline{X}_\omega$ and $\sigma_x\underline{X}^*_\omega$ and sharing 
the same imaginary part. They are linear combination of the images of these eigenstates obtained by reflecting them about axis $Oy$. 
Writing $\vec{K} = \sum_j p_j \hat{{\bf b}}_j$, these images are obtained by exchanging $p_1 \leftrightarrow p_2$ and $\delta\alpha_1 \leftrightarrow \delta\alpha_2$. 
We have confirmed this fact by verifying that the 4-dimensional eigenspace associated to four-fold degeneracies is indeed stable under reflection about axis $Oy$. 

For the cavity parameters used in the plots, the top-left panel of Fig.~\ref{fig:omega} suggests that our initial state is stable 
and superradiance cannot take place when $|U_p| \lesssim 1$. A few other stable regions appear when $|U_p|$ increases further. 
For larger values $|U_p| \gtrsim 15$, a continuous instability develops, see top-right panel of Fig.~\ref{fig:omega} where eigenvalues 
with growing positive imaginary parts are visible. 
The threshold for unstable behavior at weak pump fields does not change when the number of reciprocal lattice vectors used to compute 
and diagonalize $\mathcal{M}$ is increased. However, by expanding the momentum transfer limits and including more reciprocal lattice vectors, 
additional eigenvalues with positive imaginary parts appear in the range $|U_p| > 1$. 

A word of caution about the conclusions drawn from the linear stability analysis is necessary. Indeed, the picture may change when subsequent 
nonlinear terms are included to refine the analysis. As a matter of fact, as we have numerically checked, our initial state is still dynamically 
unstable when $|U_p|$ is chosen in between the linear instability regions visible in the top-left panel of Fig.~\ref{fig:omega}. However, when 
$|U_p|$ is chosen within the linear instability regions, a better atomic lattice structure is obtained numerically.

\section{Self-Organized Triangular and Honeycomb Lattices}
\label{numerics}

\begin{figure*}[top]
\includegraphics[width=0.72\textwidth]{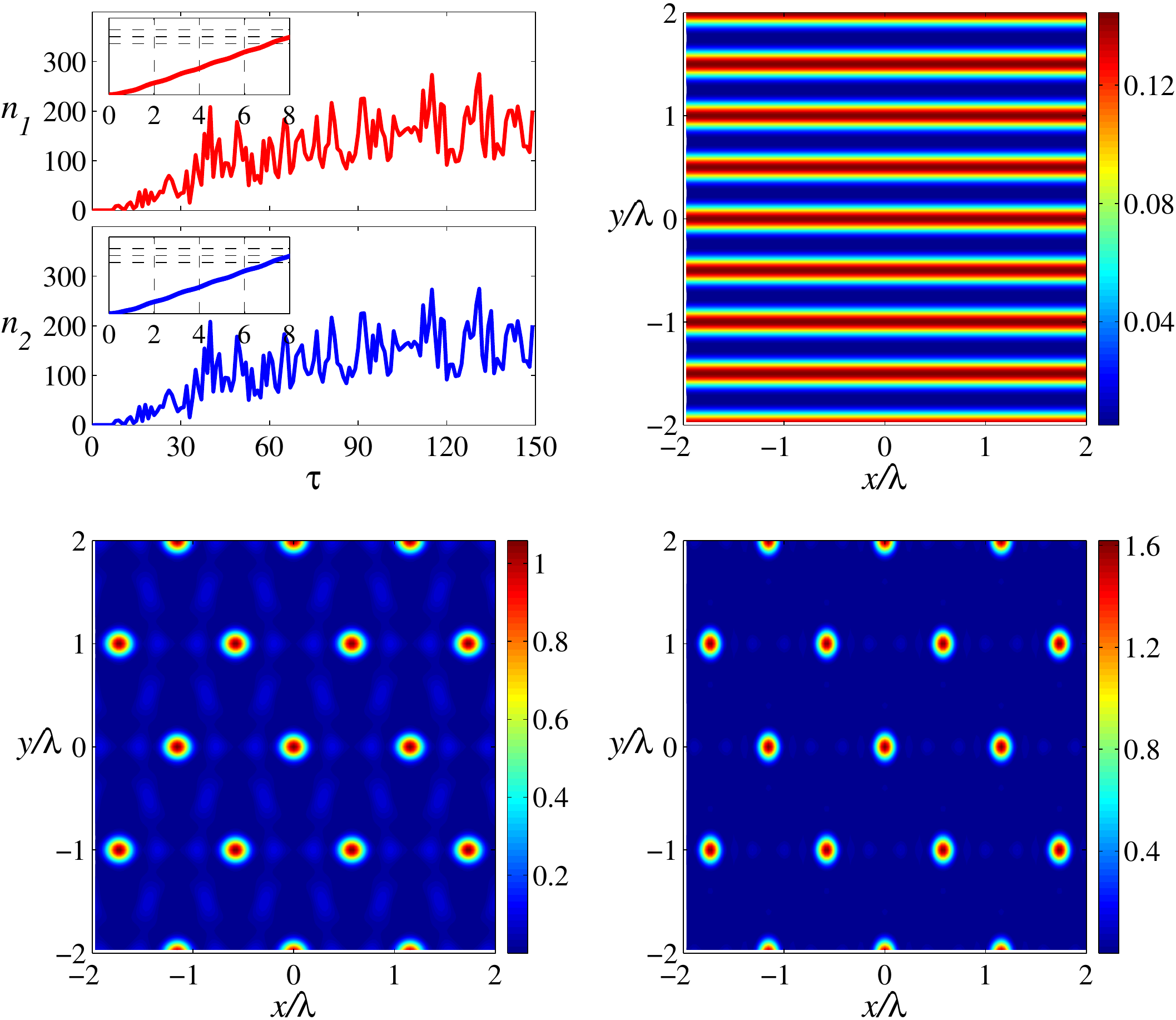}
\caption{(Color online) As a result of superradiance, an initial homogeneous atomic cloud loaded inside empty identical 
cavities self-organizes into a triangular lattice when the cavity phase is set to $\phi=0$. The pump dimensionless coupling 
strength is $U_p=-3$ and all other parameters are the same as in Fig.~\ref{fig:omega}. Atoms scatter pump photons into the cavity 
modes and, after a short while, superradiance and self-organization take place. Top-left panel: cavity photon numbers $n_j=|\alpha_j|^2$ ($j=1,2$) 
as functions of time $\tau=\omega_R t$.
The insets show the logarithms of $n_1$ and $n_2$ as a function of time and the horizontal grid lines mark the values $1$, $10$ and $100$. 
From a mathematical point of view, the symmetry of the dynamical equations and of the initial state enforces $n_1=n_2$ at all times as confirmed 
here by the numerical simulation. Top-right panel: atomic density distribution at time $\tau=2$, before superradiance takes place. The atomic 
cloud self-organizes according to the minima of the pump standing-wave potential alone since the cavity fields are almost zero. Bottom-left ($\tau=25$) 
and bottom-right ($\tau=130$) panels: atomic density distributions, and their corresponding color codes, after superradiance and self-organization 
have taken place. The atomic lattice, just like the lattice of minima of the effective potential, is clearly triangular.
\label{fig:0}}
\end{figure*}

We present now the results of our numerical simulations of the dynamical equations \eqref{eq:adot} and \eqref{eq:cdot} 
with the integrals given in Eq.~\eqref{eq:intgs2}. As mentioned in Sec.~\ref{DMA}, we only consider momenta transfer to 
atoms up to $3\hbar k$, which amounts to include thirty-seven points of the reciprocal lattice in our simulations. Starting 
from empty cavities and a homogeneous atomic cloud, the system undergoes, for appropriate laser fields strengths, a superradiant 
phase transition and subsequently self-organizes into either a triangular or a honeycomb lattice of atoms depending on the 
value of the cavity phase $\phi$. 

\subsection{Transition to the superradiant state}
\label{NtoS}

\subsubsection{Cavity phase $\phi=0$}
To check that the system does enter a superradiant phase, we monitor the dynamics of the system 
for a pump field strength larger than the lower bound found in the linear stability analysis, see Sec.~\ref{instability}. 
Our numerical simulations confirm the superradiant state for $U_p \simeq -1$. However the effective potential is not deep enough 
to support a sharp and stable lattice of atoms. This is because the pump field is too weak and cannot feed a sufficient 
number of photons inside the cavities. This problem is overcome by decreasing further the pump field strength down to $U_p=-3$ 
see Fig.~\ref{fig:0}. As shown in the top-left panel and the insets, the number of photons inside the cavities increases steadily 
and reaches $n_j \approx 10$ after a short time $\tau \simeq 8$.  
The atomic density has been computed at three different times. At small times, and before superradiance takes place ($\tau = 2$), 
the atomic cloud self-organizes according to the minima of the pump standing-wave potential alone. As superradiance takes place and 
the number of photons inside the cavities increases ($\tau=25$ and $\tau = 130$), the superposition of the pump and cavity fields 
creates a sufficiently deep potential and the atoms self-organize into the triangular lattice of potential minima, see left panel of Fig.~\ref{fig:veff}. 
It should be noted that, within the time span shown in Fig.~\ref{fig:0}, the number of cavity photons fluctuates a lot, even if it 
remains sizable, and the effective potential fluctuates too. This is because the system has not yet reached the steady state.
However these fluctuations do not alter the triangular nature of the lattice of potential minima and, in turn, the triangular nature 
of the self-organized atomic lattice.  

\begin{figure*}[top]
\includegraphics[width=0.93\textwidth]{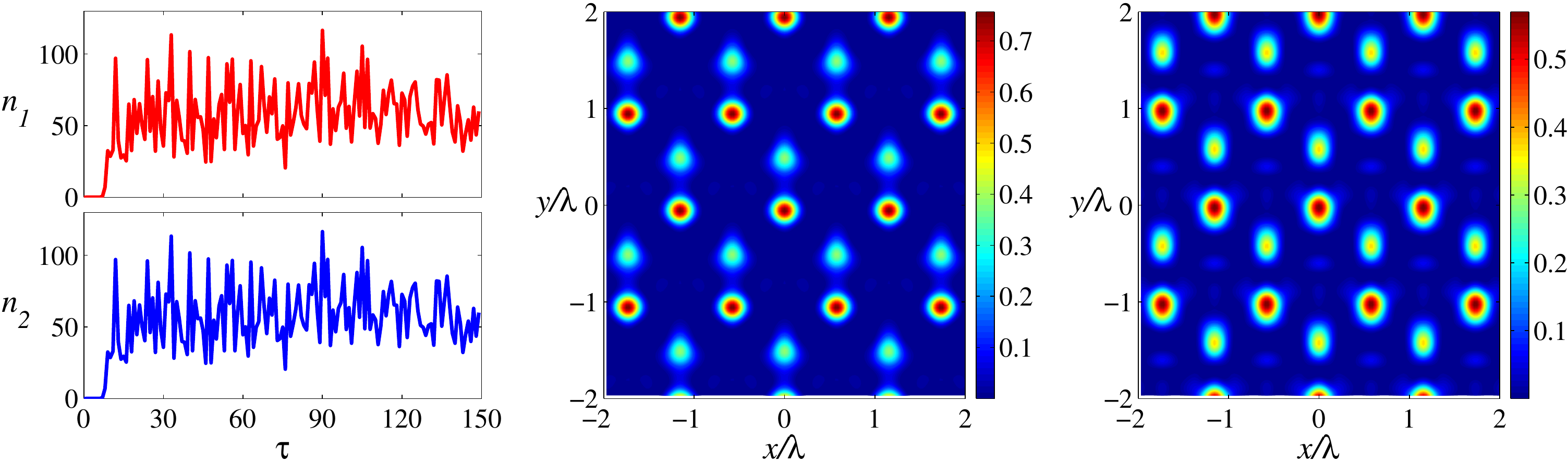}
\caption{(Color online) Same as Fig.~\ref{fig:0} but with a cavity phase set to $\phi=\pi/4$. Left panels: cavity photon numbers 
$n_j=|\alpha_j|^2$ ($j=1,2$) as functions of time $\tau=\omega_R t$. Middle ($\tau=25$) and right ($\tau=130$) panels: after 
superradiance and self-organization have taken place, an atomic honeycomb lattice with density-imbalanced consecutive sites is formed.
\label{fig:pi4th}}
\end{figure*}

\subsubsection{Cavity phase $\phi = \pi/4$}

As suggested by the results of the linear stability analysis developed in Sec.~\ref{instability}, 
superradiance takes place irrespective of the value of the cavity phase $\phi$. However the lattice structure of the potential minima is 
triangular only when $\phi = 0$, otherwise it is rather hexagonal (see Sec.~\ref{IdCav}). To check this point, we have considered the 
same initial state and we have numerically solved Eqs.~\eqref{eq:adot}, \eqref{eq:cdot} and \eqref{eq:intgs2} with $\phi=\pi/4$. In this 
case, we expect superradiance to organize atoms according to the energy-imbalanced wells of the honeycomb potential, see middle panel of 
Fig.~\ref{fig:veff} for an example of such a potential, with more atoms trapped in the deeper wells. This is indeed the behavior observed 
in Fig.~\ref{fig:pi4th} where the obtained atomic density distributions are shown at times $\tau=25$ and $\tau=130$ and display the form 
of a density-imbalanced honeycomb lattice. 

\subsubsection{Cavity phase $\phi = \pi/2$}

\begin{figure*}[top]
\includegraphics[width=0.93\textwidth]{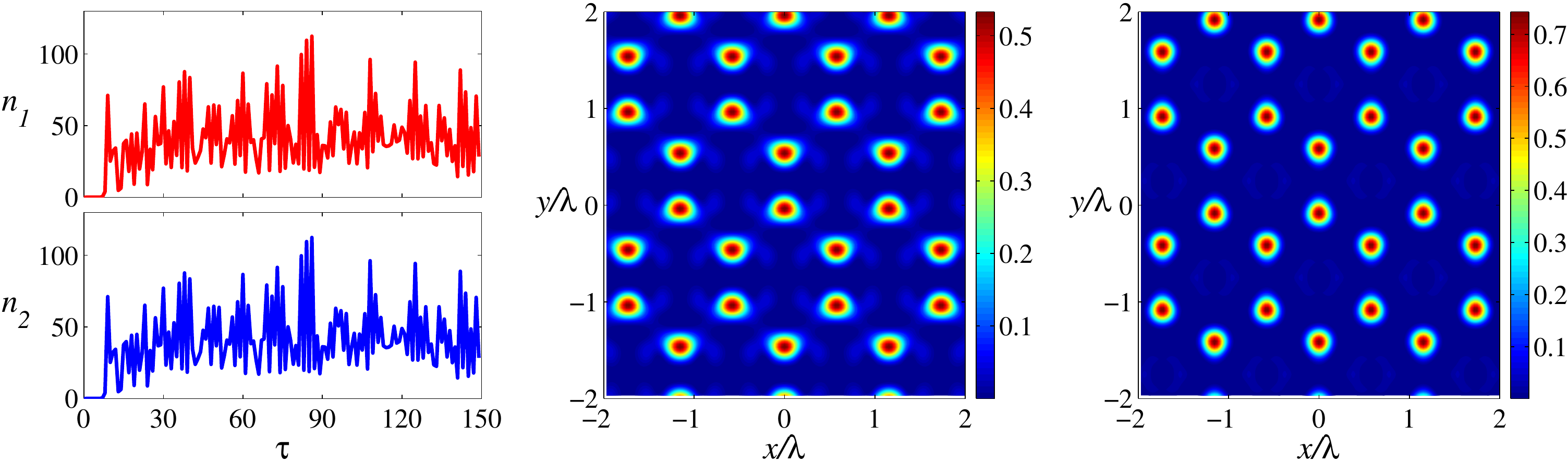}
\caption{(Color online) Same as Fig.~\ref{fig:0} but with a cavity phase set to $\phi=\pi/2$. Left panels: cavity photon numbers $n_j=|\alpha_j|^2$ ($j=1,2$) 
as functions of time $\tau=\omega_R t$. Middle ($\tau=25$) and right ($\tau=125$) panels: after superradiance and self-organization have taken place, 
an atomic honeycomb lattice with density-balanced consecutive sites is formed.
\label{fig:pihalf}}
\end{figure*}

By the same token, when $\phi=\pi/2$, the wells of the honeycomb potential have all the same depth and we expect superradiance to drive 
the atoms into a density-balanced honeycomb lattice, see right panel of Fig.~\ref{fig:veff}. This is indeed what our numerical results show, 
see Fig.~\ref{fig:pihalf} where the atomic distributions obtained at times $\tau=25$ and $\tau=125$ are given. 

One may have noticed that, for the same coupling parameters, the atomic honeycomb lattices have lower contrasts than the triangular one. 
As one can see in Fig.~\ref{fig:veff}, the honeycomb minima are already shallower than the triangular ones when the number of cavity photons 
is the same. Here the dynamics develops less cavity photons for the honeycomb lattice than for the triangular one. Moreover, for the honeycomb 
lattice, the atoms are distributed over twice many sites than for the triangular lattice. All these reasons conspire to produce less 
contrasted atomic lattices for the honeycomb structure than for the triangular one. For the parameters chosen here, we infer from Eq.~\eqref{eq:Rabi} 
that the pump and cavity fields have same order of magnitude when $n_j = U_p/U_j \approx 172$. As seen from Figs.~\ref{fig:0}-\ref{fig:pihalf}, 
this is the case for $\phi =0$ where $n_j$ fluctuates roughly around $175$. For $\phi = \pi/2$ and $\phi=\pi/4$, $n_j$ fluctuates roughly around $50$, 
or a bit less, and the pump field is larger than the cavity fields.

\subsection{Long-time stability of the different lattice structures}
\label{longtime}

\begin{figure*}[top]
\includegraphics[width=0.93\textwidth]{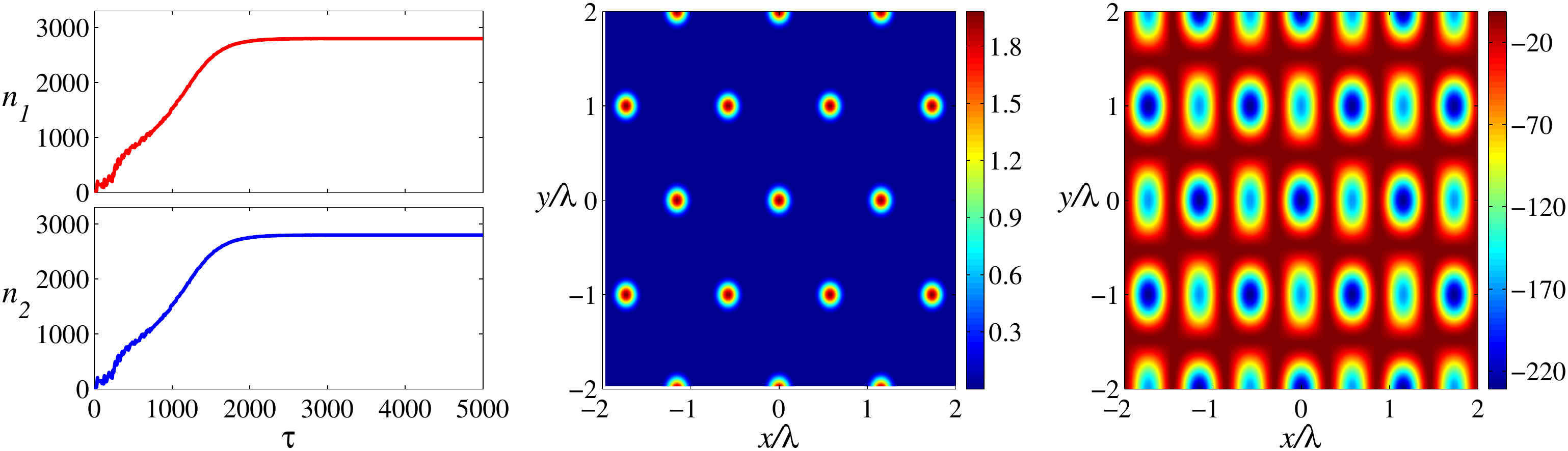}
\caption{(Color online) Long-time dynamics of the system shown in Fig.~\ref{fig:0} ($\phi=0$).
Left panel: cavity photon numbers $n_j=|\alpha_j|^2$ ($j=1,2$) as functions of time $\tau=\omega_R t$. 
The cavity steady-states are reached around a time $\tau=2000$. Middle panel: contour plot of the 
(stable) atomic density distribution obtained at time $\tau=3000$. Right panel: effective lattice 
potential $V_{\text{eff}}$ at $\tau=3000$.
\label{fig:0long}
}
\end{figure*}

As seen in Figs.~\ref{fig:0}-\ref{fig:pihalf}, the cavity photon numbers $n_j$ suffer strong temporal fluctuations. 
This means that the effective potential fluctuates too and thus, in turn, so does the atomic distribution. In fact, 
the system needs more time to reach the steady state. Fig.~\ref{fig:0long} shows the long-time evolution of the system 
when $\phi=0$ (triangular lattice). The steady state is reached around time $\tau \approx 2000$ with a number of cavity 
photons $n_1=n_2\approx 2800$. In this case the cavity fields largely dominate over the pump field and the wells of 
the effective potential organize in a rectangular array rendering the (exact) triangular symmetry less apparent as is 
shown in Fig.~\ref{fig:0long}. However, the atoms accumulate in the deepest minima of this effective potential which still 
form a nice regular triangular lattice. The contour plots in Fig.~\ref{fig:0long} show an example of the atomic density 
distribution and the effective potential at time $\tau=3000$. This distribution does not change in time once the system 
has reached the steady state. 

\begin{figure*}[top]
\includegraphics[width=0.93\textwidth]{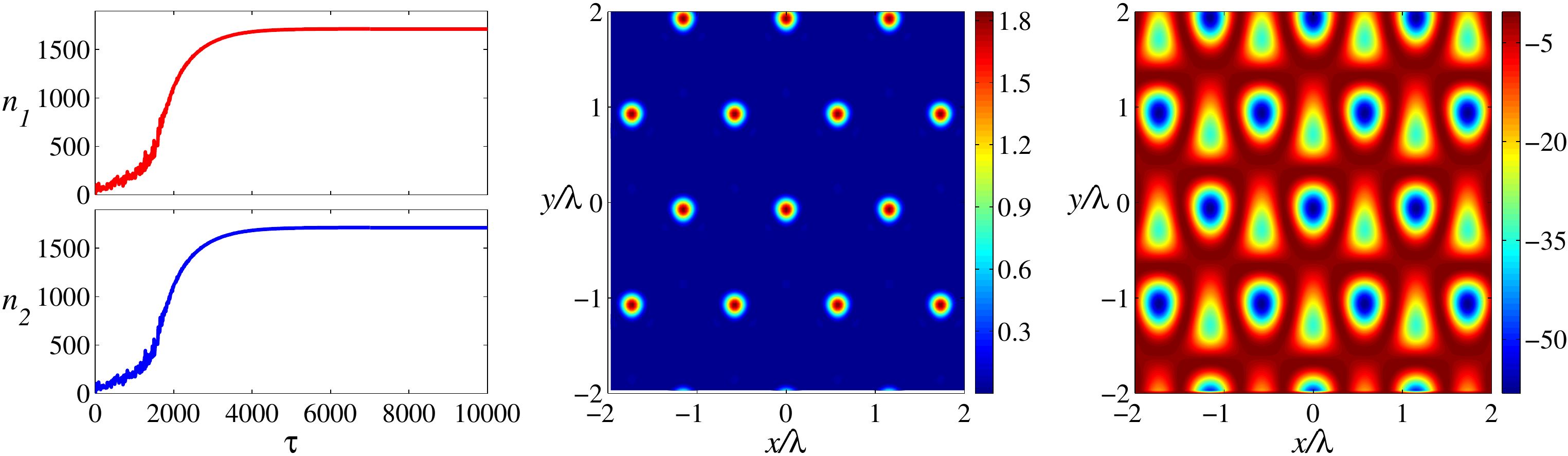}
\caption{(Color online) Long-time dynamics of the system shown in Fig.~\ref{fig:pi4th} 
($\phi=\pi/4$). Left panel: cavity photon numbers $n_j=|\alpha_j|^2$ ($j=1,2$) as functions of time $\tau=\omega_R t$. The cavity 
steady-states are reached around a time $\tau=4000$. Middle panel: contour plot of the atomic density distribution obtained at time 
$\tau=1500$. Right panel: $V_{\text{eff}}$ at $\tau=1500$. As one can see, the density-imbalanced atomic honeycomb lattice obtained at short 
times in Fig.~\ref{fig:pi4th} is not a stable structure at long times. The atoms re-organize in the deepest potential wells of the 
effective potential shown in the right panel and form a stable triangular lattice.
\label{fig:pi4thlong}}
\end{figure*}

When $0<\phi<\pi/2$, as we have seen, atomic honeycomb lattices with density-imbalanced consecutive sites are synthesized. 
However these structures are not stable in the long-time limit and atoms re-organize themselves into the deepest potential 
wells which form a triangular lattice. As an example, Fig.~\ref{fig:pi4thlong} shows the atomic density distribution and the 
effective potential obtained at time $\tau = 1500$ for $\phi=\pi/4$. As clearly seen the density-imbalanced atomic honeycomb 
lattice found at early times, see Fig.~\ref{fig:pi4th}, has disappeared in favor of the triangular atomic sub-lattice of deepest wells. 

\begin{figure}
\includegraphics[width=0.4\textwidth]{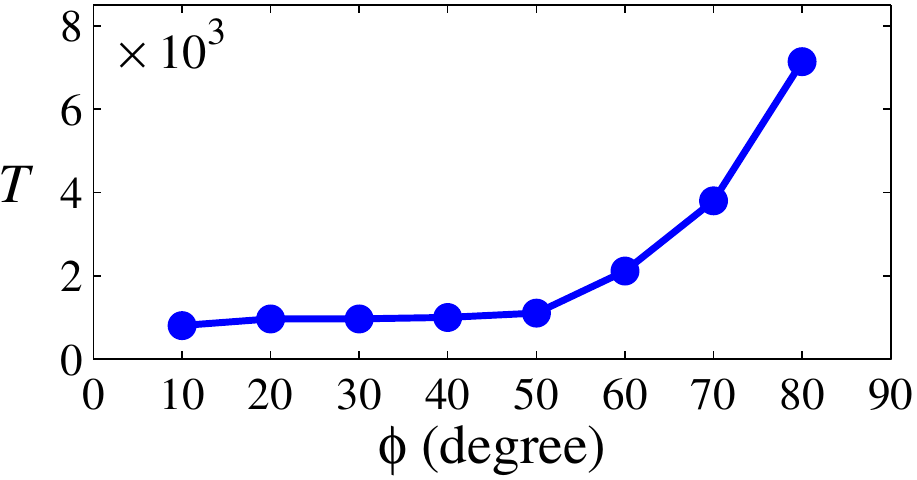}
\caption{Dimensionless latency time $T$ against the cavity phase $\phi$. After $T$, atoms have re-organized into the deepest wells 
of the energy-imbalanced effective potential obtained for $\phi\in \,]0,\pi/2[$ and exhibit a triangular lattice structure. This 
time $T$ increases when the energy mismatch between the principal and secondary potential minima decreases, i.e. when $\phi$ increases. 
It diverges when $\phi \to \pi/2$. The atomic honeycomb lattice obtained at $\phi=\pi/2$ is a stable structure.
\label{fig:T}}
\end{figure}

These density-imbalanced atomic honeycomb lattices destabilize in favor of the triangular lattice after a certain latency time $T$. 
After the time $T$, there is no visible occupation of the shallower lattice sites \footnote{In our plots, this happens when the 
maximum of the atomic density at the deeper lattice wells reaches the value $1.8$ approximately.}. The atomic lattice is purely 
triangular and becomes the stable structure in the long-time regime. The latency time $T$ depends on the potential energy difference 
between consecutive sites. Since this potential mismatch decreases when $\phi$ increases, $T$ gets longer when $\phi \to \pi/2$, 
see Fig.~\ref{fig:T}. Note however that the cavity photon numbers take a time longer than $T$ to reach their steady-state. 

\begin{figure*}[top]
\includegraphics[width=0.93\textwidth]{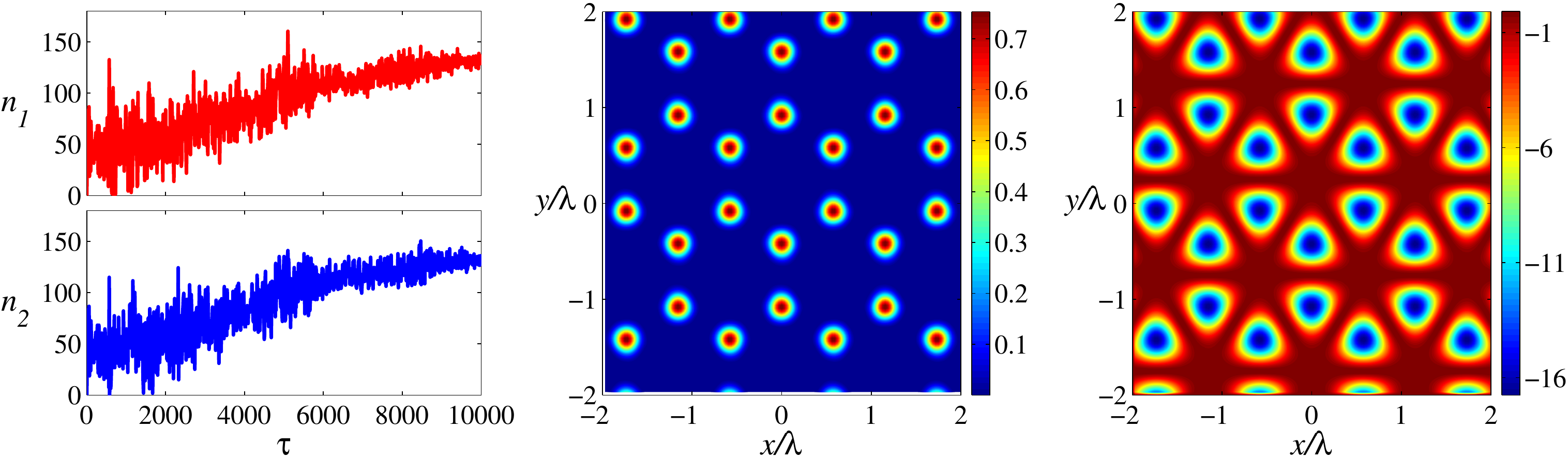}
\caption{(Color online) Long-time dynamics of the system described in Fig.~\ref{fig:pihalf} ($\phi=\pi/2$). 
Left panel: cavity photon numbers $n_j=|\alpha_j|^2$ ($j=1,2$) as functions of time $\tau=\omega_R t$. 
As one can see, the cavity photon numbers fluctuate a lot and, in turn, so does the atomic honeycomb lattice. 
At the longest time $\tau=10^4$ of our numerical simulation, the system has not yet fully reached its steady-state. 
Middle panel: corresponding contour plot of the atomic density. Right panel: effective lattice potential $V_{\text{eff}}$ 
obtained at this longest time. The atomic honeycomb lattice is a stable structure but it takes much longer time 
to reach the steady-state.
\label{fig:pihalflong}}
\end{figure*}

When $\phi=\pi/2$, the effective potential is hexagonal with perfectly energy-balanced minima. We thus expect a stable 
density-balanced atomic honeycomb lattice to emerge from the self-organization process. As one can see from Fig.~\ref{fig:pihalflong}, 
this is indeed the case, but the cavity photon numbers now reach their steady-state after a much longer time. They even fluctuate a 
lot during their temporal evolution and induce in turn fluctuations of the atomic lattice. These atomic density fluctuations get 
strongly reduced after the time $\tau=6000$ but, even at time $\tau=10^4$, the system has not yet fully reached its steady-state. 
As one may have noticed, our numerical results produce different cavity photon numbers $n_1\neq n_2$ even if their average behavior 
is the same and their asymptotic values are equal. This asymmetry is due to the sensitivity of this system to numerical errors. 
However, the difference in $n_1$ and $n_2$ approaches zero as the system reaches its steady state. 

\subsection{Driving atoms from one lattice structure to another}
\label{switch}

So far, we have considered the dynamical behavior of the cavity-atom system for a fixed pump strength, chosen above the 
threshold for superradiance, and we have studied the self-organization of atoms into triangular or hexagonal lattice 
structures for different cavity phases. We now investigate the possibility of driving atoms from one lattice structure 
to another by changing in time both the pump strength and the cavity phase. The protocol we explore is the following. 
The pump strength increases linearly in time from $0$ to some final value $U_f$ over a certain time window $3\Delta\tau$. 
Meanwhile, the cavity phase is set at $\phi_i$ initially. After superradiance takes place and enough number of photons 
are produced inside the cavities, the phase is then changed linearly in a time interval of $\Delta\tau$ to its final value 
$\phi_f$. Finally, the cavity phase keeps this target value for the rest of the time sequence.

We present below the 
time evolution of the cavity photon numbers and the atomic distribution obtained numerically for this experimental protocol 
for 3 possible cavity phase values: $\phi=0$, $\phi=\pi/2$ and some intermediate value that we choose to be $\phi = \pi/4$.

\subsubsection{Switching the cavity phase between $0$ and $\pi/4$}
\label{tri-hon1}

\begin{figure*}[top]
\includegraphics[width=0.66\textwidth]{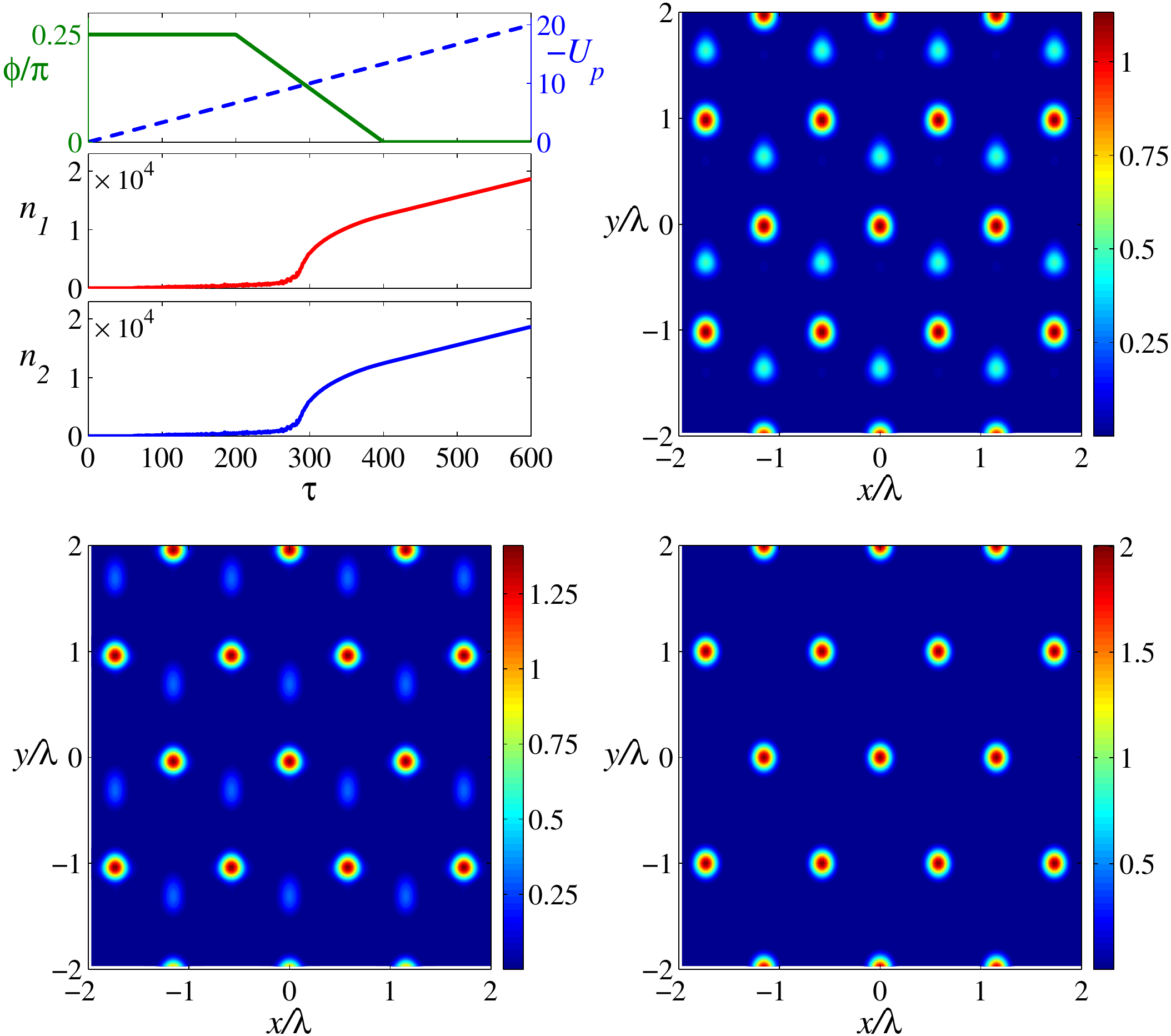}
\caption{(Color online) Transition from the density-imbalanced honeycomb atomic lattice to the triangular atomic lattice. 
Top-left panel: pump and cavity phase temporal sequences and time evolution of the cavity photon numbers. The cavity phase changes 
linearly from $\phi_i=\pi/4$ to $\phi_f=0$ over the time interval $\Delta\tau=200$. Meanwhile the pump strength $U_p$ decreases 
from 0 to $U_f=-20$ over the time window $3\Delta\tau=600$. 
The other parameters are the same as in Fig.~\ref{fig:omega}. Contour plots show the atomic density obtained at time $\tau = \omega_Rt=200$, 
before the cavity phase starts to decrease (top-right panel), at time $\tau=260$, just before the triangular lattice become the dominant 
structure (bottom-left panel), and at time $\tau=400$, when the cavity phase has reached its target value $\phi_f=0$ (bottom-right panel). 
The triangular lattice becomes the prominent structure around $\tau\approx 280$, well before the cavity phase reaches the target value $0$. 
At this time, the cavity photon numbers rise considerably and then increase linearly in time like the pump strength for the rest of the time sequence.
\label{fig:pi4h-0}}
\end{figure*}

When the system starts with $\phi_i=0$, the atoms organize in a triangular lattice. By switching the cavity phase to $\phi_f=\pi/4$, 
a new optical potential is produced featuring additional shallower wells. However we know from Sec.~\ref{longtime} that when $0<\phi<\pi/2$, 
the stable structure at long times is the triangular lattice formed by atoms occupying the deepest potential wells. Since atoms have already 
self-organized in a triangular lattice, changing the cavity phase from $0$ to $\pi/4$ will not affect the atomic distribution which thus remains triangular. 
On the other hand, when the system starts instead with $\phi_i=\pi/4$ and the cavity phase is changed to $\phi_f=0$, the transition to 
the triangular lattice is inevitable and is illustrated in Fig.~\ref{fig:pi4h-0}. 
As the cavity phase decreases, the secondary minima of the effective potential obtained for $\phi_i=\pi/4$ become shallower and shallower 
and are occupied by less and less atoms. 
The triangular lattice becomes the prominent structure around $\tau\approx 280$, well before $\phi$ 
reaches the target value $0$, which triggers a considerable rise of the cavity photon numbers.

\subsubsection{Switching the cavity phase between $0$ and $\pi/2$}
\label{tri-hon2}

\begin{figure*}
\includegraphics[width=0.66\textwidth]{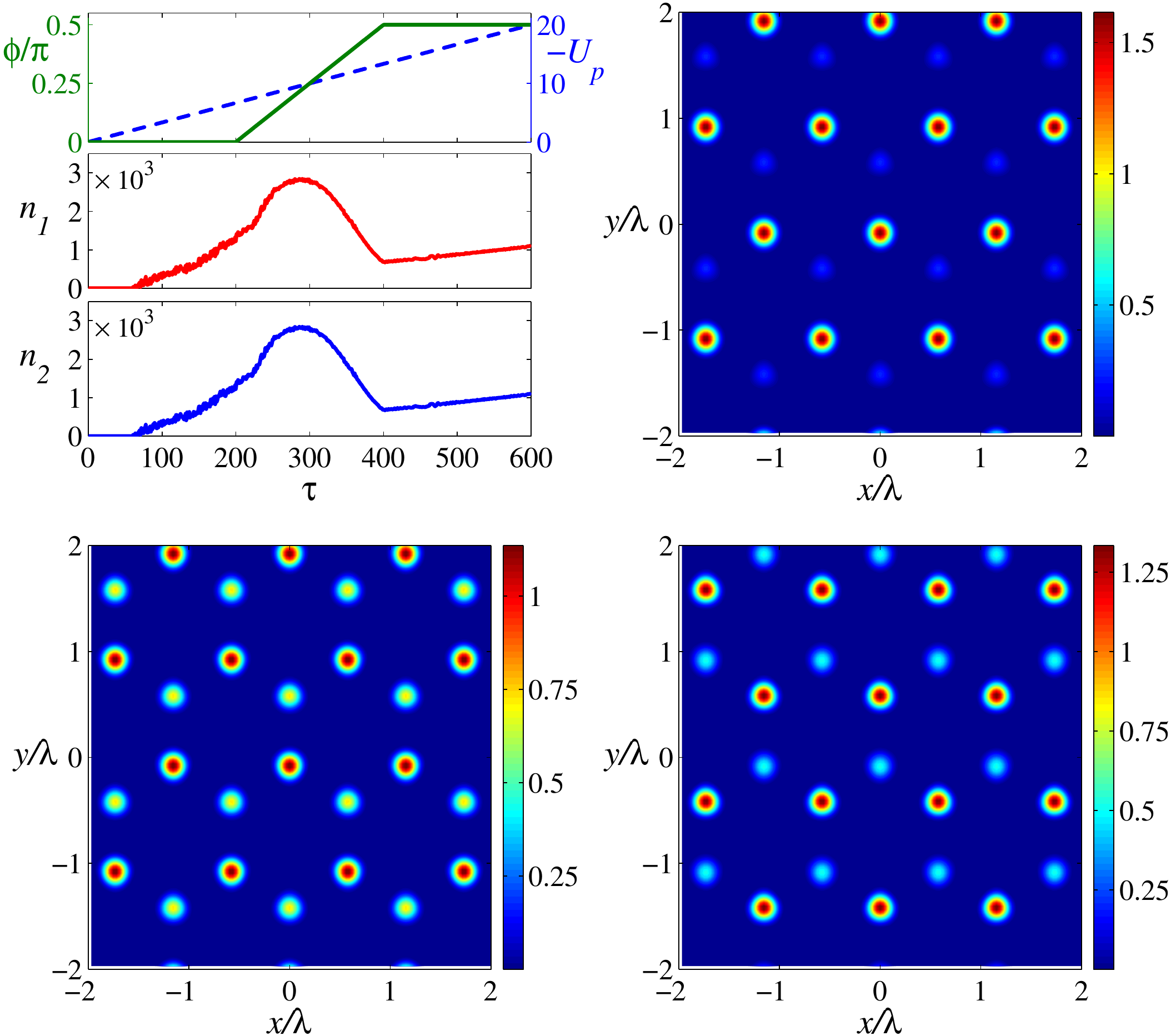}
\caption{(Color online) Transition from the triangular atomic lattice to the density-balanced atomic honeycomb lattice. 
Top-left panel: pump and cavity phase temporal sequences and time evolution of the cavity photon numbers. The cavity phase changes linearly from 
$\phi_i=0$ (triangular lattice) to $\phi_f=\pi/2$ (energy-balanced honeycomb lattice) over the time interval $\Delta\tau=200$. Meanwhile the pump 
strength $U_p$ decreases from 0 to $U_f=-20$ over the time window $3\Delta\tau=600$. The other parameters are the same as in Fig.~\ref{fig:omega}. 
When the cavity phase departs from $0$, secondary minima start to grow and the effective potential becomes an energy-imbalanced honeycomb potential. 
However, the atomic density keeps its triangular lattice structure where atoms are sitting at the deeper minima of the effective potential. when 
the cavity phase reaches $\pi/2$ then the atomic density starts oscillating in time between the two triangular sub-lattices of the honeycomb lattice 
which now have the same depth. Top-right panel shows the atomic density at time $\tau=400$ right at the time phase reaches $\pi/2$. Bottom-left panel: 
atomic density distribution at time $\tau=402$. The initial triangular lattice is still dominant. Bottom-right panel: atomic density 
distribution at time $\tau=404$. The atomic distribution now lives mainly on the other triangular sub-lattice of the honeycomb lattice.
\label{fig:0-pihalf}} 
\end{figure*}

In this case, both the triangular and density-balanced honeycomb lattices are stable structures of the static situation. 
Fig.~\ref{fig:0-pihalf} shows what happens when we start from the triangular lattice ($\phi_i=0$) and gradually increase 
the cavity phase to $\phi_f=\pi/2$. As one expects from the previous case, the atomic lattice structure remains triangular 
as long as the cavity phase has not reached its target value, $0<\phi(\tau)<\pi/2$. Interestingly enough, once the cavity 
phase has reached $\phi_f=\pi/2$, the atomic density distribution starts to oscillate between the wells of the two triangular 
sub-lattices of the energy-balanced effective potential, taking the form of density-imbalanced atomic honeycomb lattices in between. 
These two triangular sub-lattices are simply shifted by $-(\vec{a}_1+\vec{a}_2)/3$. As a net result, the atomic density looks like 
"blinking" between two shifted triangular lattices, a sign of a bi-stable behavior. This blinking does not stop if the pump strength 
is kept fixed, rather than linearly increasing, after the cavity phase has reached $\phi_f=\pi/2$. It even speeds up if the pump 
strength is ramped up faster. The bottom panels in Fig.~\ref{fig:0-pihalf} display the atomic densities at times $\tau=402$ and 
$\tau=404$, showing that, for the experimental parameters chosen, the oscillation period  between the two triangular lattices is 
comparable to the pump recoil time $\omega_R^{-1}$.
It is worth recalling that, once the cavity phase is set to $\pi/2$, the minima of the effective potential have same depth and display 
a honeycomb structure, irrespective of the cavity photon numbers (as long as they are not zero). This case should not thus be confused 
with the previous case ($0<\phi<\pi/2$) where the lattice potential has minima with different depths. Our present results then show that, 
for time-driven parameters, the density-balanced honeycomb structure is a transient state between two stable triangular structures. 

\begin{figure*}[top]
\includegraphics[width=0.66\textwidth]{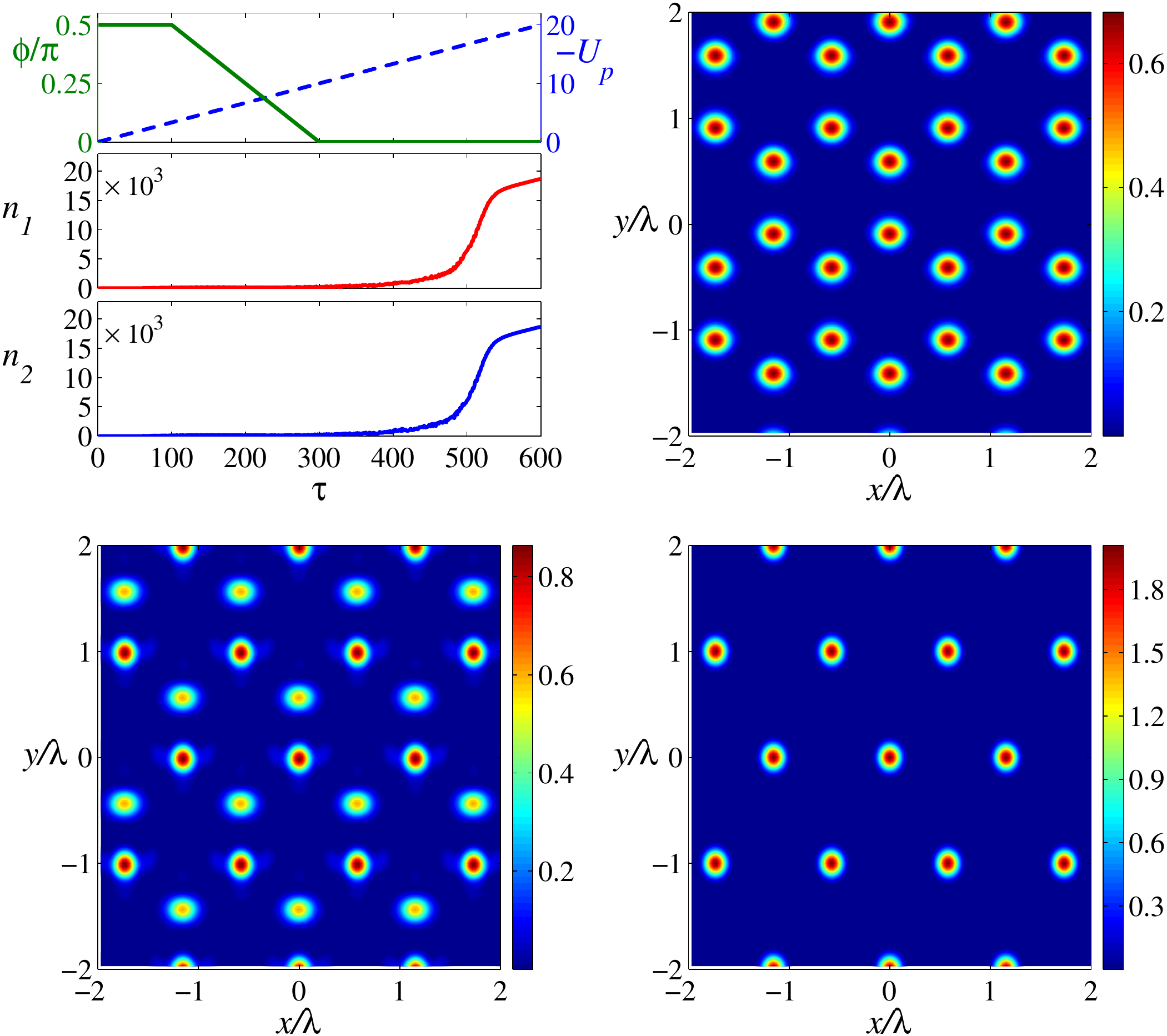}
\caption{(Color online) Transition from the density-balanced atomic honeycomb lattice to the triangular atomic lattice. 
Top-left panel: pump and cavity phase temporal sequences and time evolution of the cavity photon numbers. The cavity 
phase changes linearly from $\phi_i=\pi/2$ to $\phi_f=0$ over the time interval $\Delta\tau=200$. Meanwhile the pump 
strength $U_p$ decreases from 0 to $U_f=-20$ over the time window $3\Delta\tau=600$. The other parameters are the same 
as in Fig.~\ref{fig:omega}. Top-right panel: atomic density distribution at time $\tau=100$, forming the expected 
density-balanced honeycomb lattice. As the cavity phase departs from $\pi/2$, consecutive potential wells start to have 
different depths. As long as the potential mismatch is weak enough, the density-balanced honeycomb lattice resists the 
phase change but, with increasing potential mismatch, finally gives in and becomes a fading density-imbalanced honeycomb 
lattice around $\tau=250$ (bottom-left panel). When the cavity phase becomes even smaller, before it reaches $\phi=0$ at 
$\tau=300$, the atoms re-organize into the stable triangular structure. Bottom-right panel: atomic density distribution 
at time $\tau=600$. 
\label{fig:pihalf-0}}
\end{figure*}

Fig.~\ref{fig:pihalf-0} shows what trivially happens for the reverse process, i. e. when we start with $\phi_i=\pi/2$ and decrease 
gradually the cavity phase to $\phi_f=0$. Once the cavity phase departs from $\pi/2$, the effective potential has minima with 
different depths. As long as the potential mismatch is weak enough, the initial density-balanced honeycomb structure resists but, 
with increasing potential mismatch, finally gives in and a density-imbalanced honeycomb lattice is formed around $\tau=250$. Then, 
as the cavity phase decreases further, the depth of the secondary minima, as well as the number of atoms sitting on them, decreases 
rapidly. Before the cavity phase finally reaches $\phi=0$ at $\tau=300$, the atoms  choose the deeper triangular sub-lattice which 
becomes the stable structure as expected.

\subsubsection{Switching the cavity phase between $\pi/4$ and $\pi/2$}
\label{hon1-hon2}
\begin{figure*}
\includegraphics[width=0.66\textwidth]{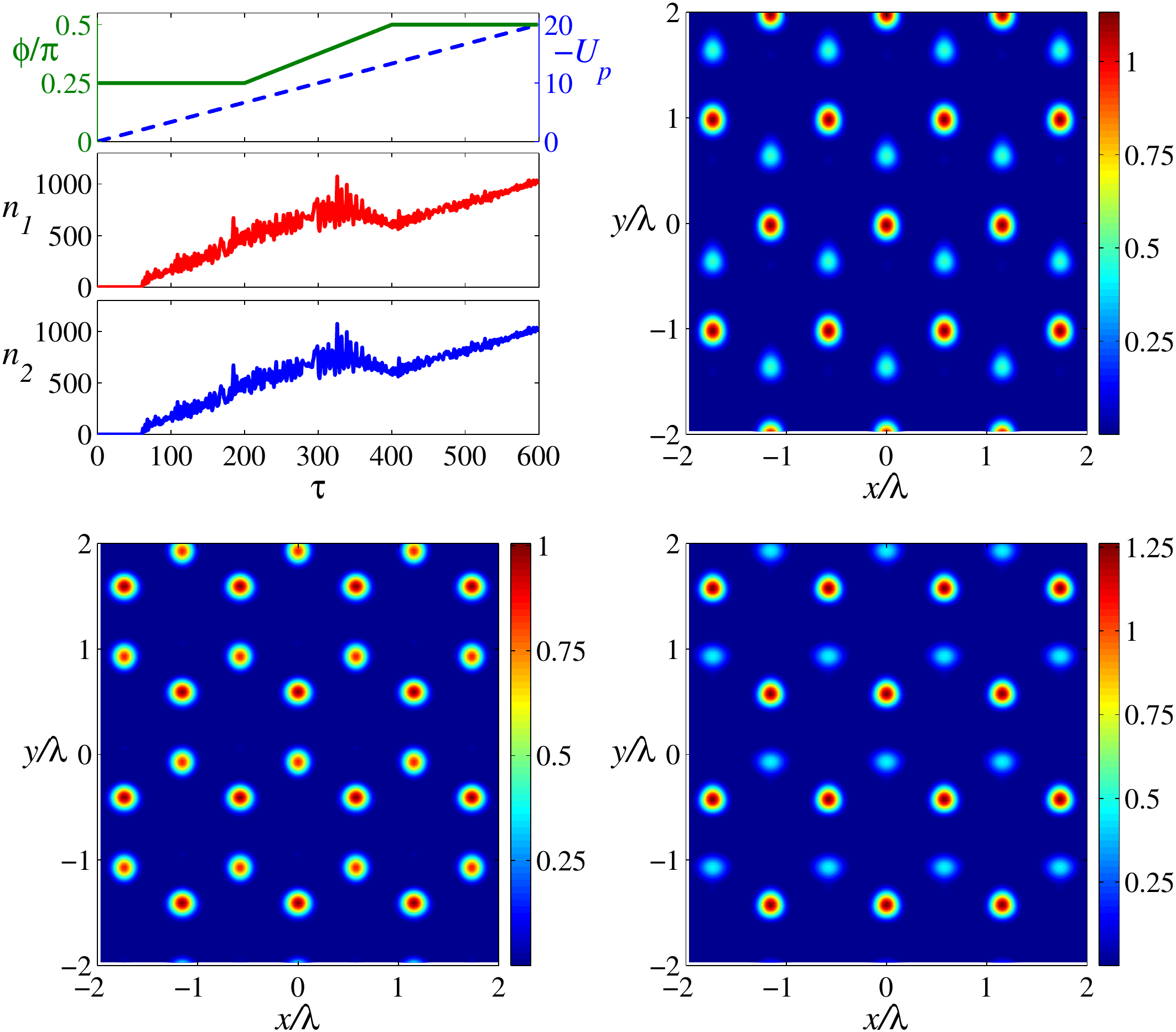}
\caption{(Color online) Transition from the density-imbalanced to the density-balanced atomic honeycomb lattice. 
Top-left panel: pump and cavity phase temporal sequences and time evolution of the cavity photon numbers. 
The cavity phase increases linearly from $\phi_i=\pi/4$ to $\phi_f=\pi/2$ over the time interval $\Delta\tau=200$. 
Meanwhile the pump strength $U_p$ decreases from 0 to $U_f=-20$ over the time window $3\Delta\tau=600$. The other 
parameters are the same as in Fig.~\ref{fig:omega}. 
Top-right panel: atomic density at time $\tau=200$ where the cavity phase starts to increase. Later on, as the cavity 
phase increases, the shallower wells become deeper, however the atoms in deeper wells do not move to the shallower wells 
until phase becomes exactly $\pi/2$ and all sites have same depth. 
Bottom-left panel: atomic density at time $\tau=400$ when the cavity phase reaches $\pi/2$ one gets an almost 
density-balanced atomic honeycomb lattice. From this time on, and for the rest of the time sequence, the atomic 
density starts oscillating fast in time between the two stable triangular sub-lattices. Bottom-right panel: 
hinting at this oscillating behavior, the atomic density at time $\tau=401$ 
\label{fig:pi4th-pihalf}} 
\end{figure*}
\begin{figure*}
\includegraphics[width=0.66\textwidth]{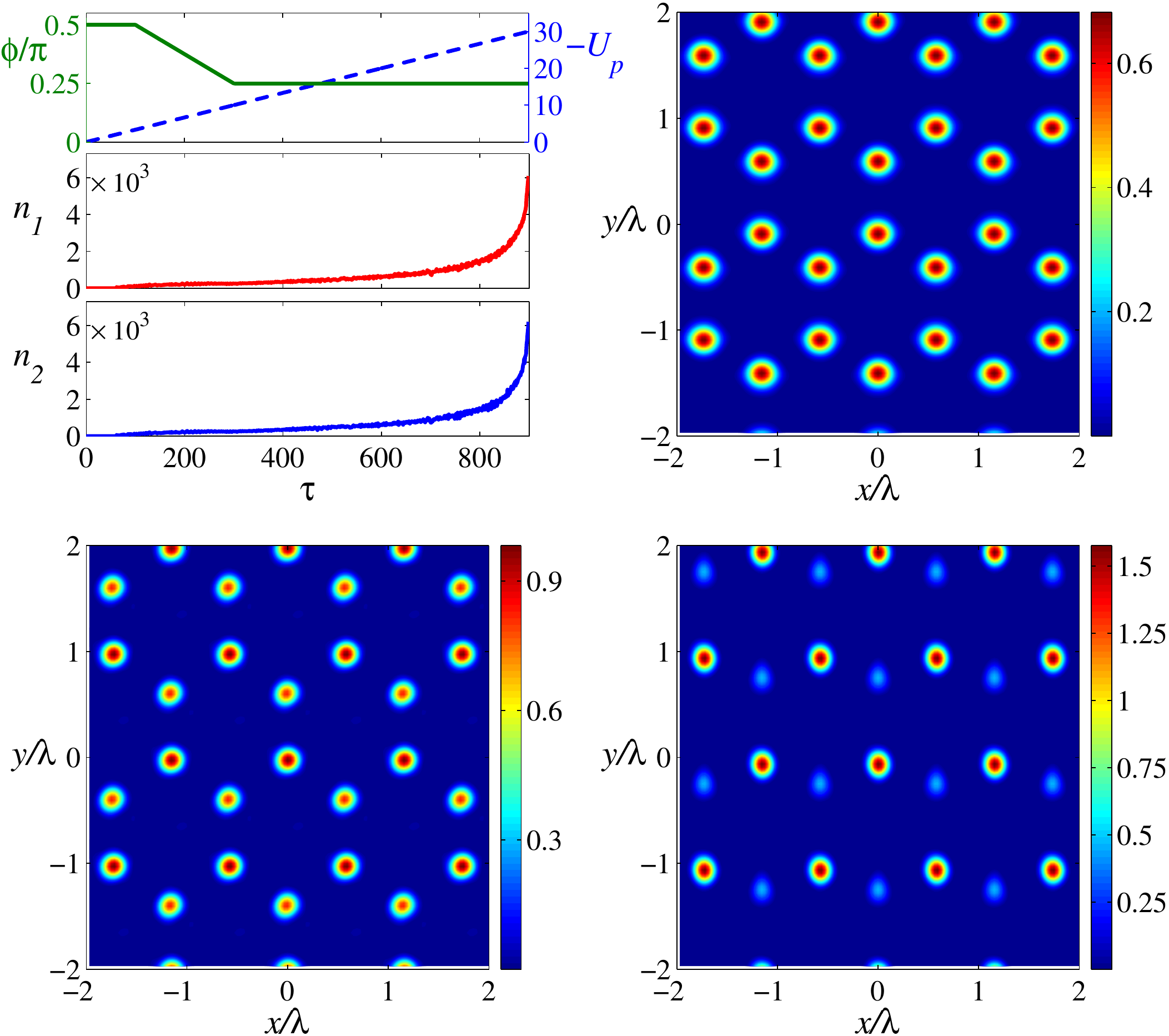}
\caption{(Color online)
Transition from the density-balanced to the density-imbalanced atomic honeycomb lattice. Top-left panel: pump and cavity 
phase temporal sequences and time evolution of the cavity photon numbers. The cavity phase decreases linearly from 
$\phi_i=\pi/2$ to $\phi_f=\pi/4$ over the time interval $\Delta\tau=200$. Meanwhile the pump strength $U_p$ decreases 
from 0 to $U_f=-30$ over the time window $4.5\Delta\tau=900$ (same rate as previous cases). The other parameters are the 
same as in Fig.~\ref{fig:omega}. Top-right panel: atomic density at time $\tau=100$ when the cavity phase is $\pi/2$ and 
one gets a density-balanced atomic honeycomb lattice. When the phase departs from $\pi/2$, similar to what happens in 
Fig.~\ref{fig:pihalf-0}, the initial density-balanced honeycomb lattice resists destabilization, even though the lattice 
potential is imbalanced, but finally gives in around $\tau\approx 600$, well after the cavity phase has reached its final 
value $\phi= \pi/4$. Bottom-left panel: atomic distribution at $\tau=650$. Bottom-right panel: atomic distribution at $\tau=900$. 
The atomic lattice transits to the triangular lattice at long times through a density-imbalanced honeycomb lattice.
\label{fig:pihalf-pi4th}} 
\end{figure*}

We address now the last case, namely the transition between the two possible, density-balanced or density-imbalanced, atomic honeycomb lattices. 
We first consider switching the cavity phase from $\phi_i=\pi/4$ to $\phi_f=\pi/2$. In this case, some of the atoms are sitting in the shallower 
potential wells of the effective honeycomb potential while the majority of them are hosted in the deeper ones. By increasing the cavity phase, 
the shallower sites become deeper. However, the atoms in the deeper wells tend to stay where they are (see Sec.~\ref{tri-hon1}) until the cavity 
phase becomes exactly $\pi/2$ and all potential wells have same depth. At this point, the atomic lattice is hexagonal with almost density-balanced 
sites. Then, for the rest of the time sequence, atoms start oscillating in time between the two stable triangular sub-lattices, featuring the 
"lattice blinking" already observed in the preceding Sec.~\ref{tri-hon2}. 

When the system starts with $\phi_i=\pi/2$, the initial atomic lattice is a density-balanced honeycomb lattice. By gradually decreasing the cavity phase, 
the potential wells of one of the two triangular sub-lattices become shallower. However, the initial density-balanced atomic honeycomb lattice resists the 
potential mismatch even after the cavity phase has reached its final value $\phi=\pi/4$. It gets finally destabilized around $\tau\approx 600$ and takes 
the form of a density-imbalanced honeycomb lattice, see Fig.~\ref{fig:pihalf-pi4th}. As explained in Sec.~\ref{longtime}, this structure is not stable in 
the long-time limit and destabilizes subsequently into a triangular lattice as the atoms in the shallower sites move to the deeper sites.

\section{Conclusion}
\label{sum}
In this work we have proposed to load a two-dimensional cloud of non-interacting cold bosonic atoms inside two identical 
initially-empty optical cavities with an angle of $2\pi/3$ between their axes. The atoms are driven by an external laser 
field. We have given the Hamiltonian of this hybrid system in the dispersive regime and we have derived the corresponding 
dynamical equations in the mean-field regime. The coherent superposition of the cavity and pump fields creates a dynamical 
effective lattice potential with a triangular Bravais structure in which the atoms move. As a result of superradiance, the 
atoms self-organize into a triangular or a honeycomb lattice inside the cavities, the nature of the lattice depending on 
the relative phase between the pump and the cavity fields. 
Using the symmetry properties of the effective potential, we have derived the dynamical equations in reciprocal space and 
we have investigated the condition required for superradiance to take place. Linear response theory shows that superradiance 
takes place irrespective of the relative phase between the cavity and pump fields. This is confirmed by our numerical results: 
atoms self-organize into a triangular lattice when pump and cavity fields are in phase, into a density-balanced honeycomb 
lattice when the fields are in quadrature and into a density-imbalanced honeycomb lattice in between. The stable atomic structures 
in the long-time limit are the triangular and density-balanced honeycomb lattices. The density-imbalanced honeycomb lattice 
only survives for a limited amount of time which becomes longer and longer as the cavity and pump fields approach quadrature. 
In the end, atoms redistribute equally into the triangular sub-lattice made of the deepest potential wells. 
We have also studied the transition between these different lattice structures when the relative phase between the fields is 
dynamically changed and confirmed the stability of the triangular and density-balanced honeycomb structures.
A natural extension of this work would include the study of the survival time of the density-imbalanced honeycomb lattices, 
the (possibly chaotic) dynamics leading to superradiance when the fields are in quadrature and, importantly, the effect of 
atomic interactions on the self-organization process.

\acknowledgments
Sh. Safaei would like to gratefully acknowledge inspiring discussions with \"Ozgur Esat M\"ustecapl{\i}o\u{g}lu. 
The Centre for Quantum Technologies is a Research Centre of Excellence funded by the Ministry of Education and 
National Research Foundation of Singapore.


\end{document}